\documentclass[a4paper,fleqn,usenatbib]{mnras}

\usepackage[T1]{fontenc}
\usepackage{ae,aecompl}

\usepackage{graphicx}	
\usepackage{amsmath}	
\usepackage{amssymb}	

\newcommand{\beq}{\begin{equation}}
\newcommand{\eeq}{\end{equation}}
\newcommand{\beqn}{\begin{eqnarray}}
\newcommand{\eeqn}{\end{eqnarray}}
\newcommand{\grim}{{\tt grim}}
\newcommand{\rS}{{G M/c^2}}
\newcommand{\tS}{{G M/c^3}}
\newcommand{\<}{{\langle}}
\renewcommand{\>}{{\rangle}}

\title[Accretion Discs in Extended Magnetohydrodynamics]
{Evolution of Accretion Discs around a Kerr Black Hole using Extended Magnetohydrodynamics}

\author[F. Foucart et al.]{
Francois Foucart$^{1}$\thanks{E-mail: fvfoucart@lbl.gov},
Mani Chandra$^{2}$,
Charles F. Gammie$^{2}$,
Eliot Quataert$^{3}$\\
$^{1}$Lawrence Berkeley National Laboratory, 1 Cyclotron Rd, Berkeley, CA 94720, USA; Einstein Fellow\\
$^{2}$Department of Astronomy and Department of Physics, University of Illinois, 1002 West Green Street, Urbana, IL 61801\\
$^{3}$Department of Astronomy and Theoretical Astrophysics Center, University of California, Berkeley, CA, 9472
}

\date{Accepted XXX. Received YYY; in original form ZZZ}

\pubyear{2015}

\begin{document}
\label{firstpage}
\pagerange{\pageref{firstpage}--\pageref{lastpage}}
\maketitle

\begin{abstract}
Black holes accreting well below the Eddington rate are believed to have geometrically thick, optically thin, rotationally supported accretion discs in which the Coulomb mean free path is large compared to $GM/c^2$. In such an environment, the disc evolution may differ significantly from ideal magnetohydrodynamic predictions. We present non-ideal global axisymmetric simulations of geometrically thick discs around a rotating black hole.  The simulations are carried out using a new code {\tt grim}, which evolves a covariant extended magnetohydrodynamics model derived by treating non-ideal effects as a perturbation of ideal magnetohydrodynamics. Non-ideal effects are modeled through heat conduction along magnetic field lines, and a difference between the pressure parallel and perpendicular to the field lines. The model relies on an effective collisionality in the disc from wave-particle scattering and velocity-space (mirror and firehose) instabilities. We find that the pressure anisotropy grows to match the magnetic pressure, at which point it saturates due to the mirror instability.  The pressure anisotropy produces outward angular momentum transport with a magnitude comparable to that of MHD turbulence in the disc, and a significant increase in the temperature in the wall of the jet.  We also find that, at least in our axisymmetric simulations, conduction has a small effect on the disc evolution because (1) the heat flux is constrained to be parallel to the field and the field is close to perpendicular to temperature gradients, and (2) the heat flux is choked by an increase in effective collisionality associated with the mirror instability.  
  
\end{abstract}




\section{Introduction}
\label{sec:intro}

The Event Horizon Telescope (EHT) is now reaching the sensitivity required to image with sub-horizon resolution the accretion discs around Sgr~A$^*$ and around the supermassive black hole at the center of the M87 galaxy~\citep{Doeleman2009}. These images may help constrain the properties of black holes and of accretion flows close to a black hole horizon, where general relativistic effects are expected to be important.

Most supermassive black holes, including both black holes which can be resolved by the EHT, are accreting well below the Eddington rate~\citep{Ho2009}, in a regime in which their accretion disc
is expected to be geometrically thick and optically thin. Phenomenological models of these radiatively inefficient accretion
flows (RIAF, see \citealt{Yuan2014} for a review), and numerical simulations of those discs with general relativistic codes 
evolving the equations of ideal 
magnetohydrodynamics (MHD)~\citep{Koide1999,Devilliers2003,McKinney2004}, tell us that the physical conditions within the disc are such that the Coulomb mean free path 
of both ions and electrons is much larger then the typical length scale of the disc, $\sim GM/c^2$~\citep{Mahadevan1997}, where
$M \equiv $ black hole mass, $G \equiv$ Newton's constant, and $c \equiv$ speed of light. 
The plasma in these discs is thus expected to
be collisionless. This naturally raises questions about the validity of modeling the discs as ideal fluids. 

At first glance, studies of these collisionless plasmas should require the evolution of the distribution function of both ions and electrons, a problem that, at present, is too computationally expensive for global 3D simulations. However, particle-in-cell simulations of shear flows and expanding/contracting boxes have shown that wave-particle interactions generated by velocity space instabilities increase the effective collision rate of particles~\citep{Kunz2014,Riquelme2015,Sironi2015,Hellinger2015}.  This conclusion is supported by measurements in the solar wind, which show that the plasma pressure anisotropy is bounded by the thresholds associated with linear velocity space instabilities (e.g., \citealt{Kasper2002,Hellinger2006}). Likewise, galactic cosmic rays are remarkably isotropic despite their enormous collisional mean free paths, indicating an efficient source of wave-particle scattering (e.g., \citealt{Kulsrud2004}). These theoretical and observational results motivate an approach in which non-ideal effects are treated as perturbations relative to an ideal fluid, via the inclusion of physics such as anisotropic heat and momentum transport.  This is what we shall call the {\it weakly collisional} plasma model.  Although formally of questionable applicability to the collisionless plasmas of interest, these non-ideal fluid models may nonetheless provide useful insight into the importance of non-ideal  physics for global accretion disc dynamics.

\begin{figure}
\flushleft
\includegraphics[width=1.\columnwidth]{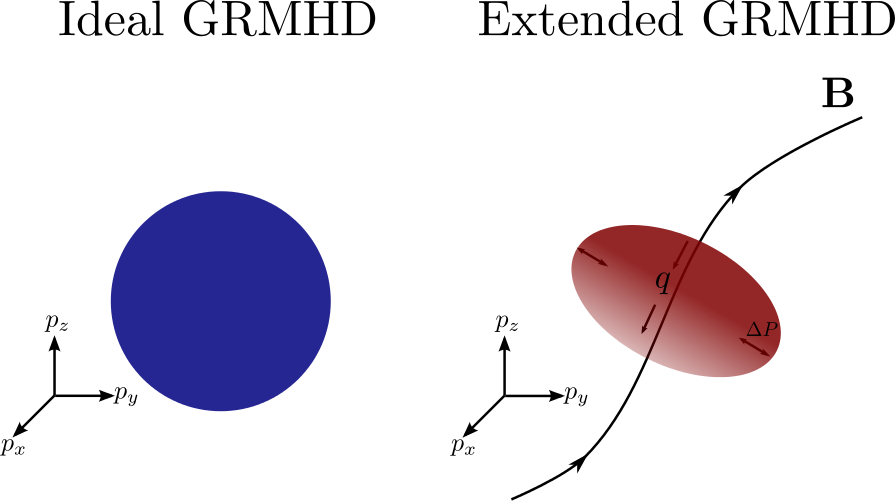}
\caption{Sketch illustrating the differences in the momentum space distribution functions between ideal MHD and extended MHD (EMHD). Ideal MHD assumes local thermodynamic equilibrium, and the distribution function at every spatial location is a thermal distribution which is uniquely specified by two thermodynamic variables, the density $\rho$ and the temperature $\Theta$. Extended MHD incorporates the effect of a pressure anisotropy $\Delta P$ and an anisotropic heat flux $q$ in the presence of a magnetic field $\mathbf{B}$, thus evolving two additional variables.}
\label{fig:cartoon}
\end{figure}

In~\cite{Chandra2015a}, hereafter Paper I, we  derived a covariant model for a weakly collisional magnetized
plasma, which we will refer to as an extended magnetohydrodynamics (EMHD) model. The model was derived by exploiting the symmetries of the distribution function in the presence of a magnetic field. Because of these symmetries, the deviations from ideal MHD manifest themselves as a heat flux flowing along magnetic field lines, and a pressure anisotropy, i.e.~a difference between the pressure parallel and perpendicular to the 
local magnetic field lines, which is directly related to the viscous stress tensor (see Fig.~\ref{fig:cartoon}). The evolution equations for the heat flux and pressure anisotropy are derived by invoking the weakly collisional approximation, building on the relativistic theory
of \cite{Israel1979}. In the non-relativistic limit, the model in Paper I reduces to \cite{Braginskii1965}'s theory of anisotropic, weakly collisional plasmas. The EMHD model, its free parameters, and the connection between those free parameters and kinetic theory are discussed in \S \ref{sec:methods}.

The EMHD model presents a number of computational challenges compared with the evolution of the standard equations of ideal MHD. The dissipative fields in the EMHD model (heat flux and pressure anisotropy) are driven by \emph{spatio-temporal} gradients of the fluid thermodynamic quantities, not just spatial gradients as in non-relativistic dissipative theories. This leads to time derivatives of thermodynamic variables as source terms in the evolution equations for the dissipative fields.  In addition, the stress tensor in the EMHD model has contributions from the heat flux and pressure anisotropy, whose \emph{spatio-temporal} gradients in turn contribute to the evolution of the momentum and energy densities. This leads to a non-linear coupling between the energy and momentum conservation equations and the evolution equations for heat flux and pressure anisotropy. This coupling voids the algorithms used for primitive variable recovery in current conservative codes. Further, the EMHD model has source terms that become stiff in certain situations (see \S \ref{sec:closure}). To deal with these issues we have developed a new code, \grim. The equations and algorithms used in \grim \ are summarized in 
\S \ref{sec:grim}. A more detailed description of the code and an extensive series of code tests will be described in
an upcoming publication (Chandra et al. 2015b, in prep., hereafter Paper II). The code and 
test results are available on the \grim~website\footnote{http://afd-illinois.github.io/grim/}.

In this paper we begin to characterize non-ideal effects in global models of sub-Eddington accretion discs 
by performing axisymmetric EMHD simulations of discs
around a rapidly rotating black hole. Through these simulations, we can model RIAFs with improved realism and assess the importance of both pressure anisotropy and heat conduction for the evolution of the disc.  To do so, we perform a series of 
simulations starting from hydrodynamical equilibrium, with a seed magnetic field and small pressure perturbations.   
The initial conditions considered here are described in \S \ref{sec:ID}, while \S \ref{sec:results} presents the simulations, and 
\S \ref{sec:conclusions} contains a summary and conclusion.
In the simulations, we implicitly assume $P_e/P_i \ll 1$ in regions where gas pressure forces are important,
 where $P_e$ and $P_i$ are the electron and ion pressures. With this assumption, the fluid can be interpreted as the ions in RIAF models, with the electrons not contributing to the overall dynamics. 

\section{Model and Numerical Methods}
\label{sec:methods}

\subsection{Extended Magnetohydrodynamics Model}
\label{sec:emhd}

We consider the evolution of the plasma surrounding a rotating black hole
accreting at a sub-Eddington rate using the single fluid EMHD model proposed in Paper
I. The EMHD model takes advantage of the expected ordering of length scales in
which the Larmor radius of particles is much smaller than both the
typical length scale of the system $GM/c^2$ and the Coulomb (or
wave-particle) mean free path. Particles are thus effectively confined
in the direction orthogonal to the magnetic field, but propagate
freely along magnetic field lines.  In the systems of interest, the
collisional mean free path is large compared to $GM/c^2$, which
suggests that a fluid theory is completely inappropriate. However,
particle-in-cell simulations have shown that wave-particle
interactions generated by velocity-space instabilities create an
effective collision rate in these systems
\citep{Kunz2014,Riquelme2015,Sironi2015,Hellinger2015}, better
motivating a model which treats non-ideal effects as a perturbation
relative to ideal MHD.  It is, however, an open question exactly how
well this approach compares to full kinetic theory
calculations.

In the limit in which the Larmor radius is small compared to the mean free
path, it is reasonable to assume that the distribution function of the
ions is symmetric with respect to rotations in momentum space around
the direction of the magnetic field.  In the non-relativistic limit (e.g., \citealt{Braginskii1965}) non-ideal effects are then modeled through heat conduction along the magnetic field lines
\beq
q_i = -\rho \chi \hat{B}_i \hat{B}_j \frac{\partial \Theta}{\partial x_j},
\eeq
and a viscous stress tensor
\beq
\Pi_{ij} = -\Delta P \left( \hat{B}_i \hat{B}_j - \frac{1}{3}\delta_{ij} \right).
\eeq 
Here we use Cartesian coordinates, $\hat{B}_i$ is a unit 3-vector parallel to the magnetic field, $\chi \equiv$ conductive diffusivity (units of length times velocity), $\Theta$ is proportional to the temperature, and $\rho \equiv$ density.   In collisional, nonrelativistic theory
\beq
\Delta P = 3\rho\nu \left(\hat{B}_i \hat{B}_j \frac{\partial  {\rm v}_j}{\partial x_i} - \frac{1}{3} \frac{\partial {\rm v}_k}{\partial x_k}\right),
\eeq
where ${\rm v}_i$ is the fluid 3-velocity and $\nu$ is the kinematic viscosity.   The viscous stress is directly related to the pressure anisotropy $\Delta P = P_\perp - P_\parallel$ between the pressure perpendicular to the magnetic field lines, $P_\perp$, and the pressure along the magnetic field lines, $P_\parallel$ (the inverse relations are $P_\parallel =P - \frac{2}{3} \Delta P$ and $P_\perp  = P+\frac{1}{3} \Delta P$); one can describe the additional energy-momentum fluxes as arising from a viscosity or equivalently a pressure anisotropy.

The \cite{Braginskii1965} model has been widely used to study laboratory and astrophysical plasmas, but generalizing it to relativistic systems is not trivial. Simply adding the relativistic equivalent of Braginskii's heat flux and viscous stress tensor to the equations of general relativistic magnetohydrodynamics leads to an ill-posed (unstable and acausal) model. This issue is well-known for isotropic heat fluxes and viscosities \citep{Hiscock1985}, and we confirmed in Paper I that the same problem arises in the anisotropic case.  A slightly more complicated model, in which the heat flux and pressure anisotropy are promoted to dependent variables relaxing to their desired value on a dynamical timescale, is needed to restore well-posedness.  We derived such an extended magnetohydrodynamics (EMHD) model for weakly collisional plasmas in Paper I, inspired by the theory for isotropic conduction and viscosity developed by \cite{Israel1979}, which was proven to be well-posed by \cite{Hiscock1983}.

\subsection{Evolution Equations}
\label{sec:eq}

We now describe the evolution equations used in our accretion disc simulations. Here and in the following we assume a background Kerr black hole spacetime with mass $M$ and dimensionless spin $a_*$.  We set $G M = c =1$; greek indices run over four coordinate components, latin indices run over the spatial components only, the metric is $g_{\mu\nu}$, and its determinant is $g$.

The ideal MHD stress-energy tensor is
\beq
T^{\mu\nu}_{\rm ideal} = (\rho + u + P + b^2)u^\mu u^\nu + (P + b^2/2) g^{\mu\nu} - b^\mu b^\nu
\eeq
where $b^\mu$ is the magnetic field 4-vector, which reduces to the ordinary magnetic field in the fluid frame, $b^2 = b^\mu b_\mu$, $\rho$ is the baryon density, $u$ is the internal energy density, and $P$ is the gas pressure. 

In the EMHD model the total stress-energy tensor is
\beq
T^{\mu \nu} = T^{\mu\nu}_{\rm ideal} + q^\mu u^\nu + q^\nu u^\mu + \Pi^{\mu\nu}
\eeq
where $q^\mu = q \hat{b}^\mu$ is the heat flux, $\hat{b}^\mu = b^\mu/b$ is the unit vector along $b^\mu$,
\beq
\Pi^{\mu\nu} = - \Delta P \left( \hat{b}^\mu \hat{b}^\nu - \frac{1}{3} h^{\mu\nu} \right)
\eeq
is the viscous stress-energy tensor, and 
\beq
h^{\mu\nu}=g^{\mu\nu}+u^\mu u^\nu
\eeq
is the projection on a hypersurface orthogonal to the 4-velocity $u^\mu$.   The model does not include bulk viscosity.  These choices for $q^\mu$ and $\Pi^{\mu\nu}$ permit energy and momentum to be transported along but not across magnetic field lines.

The evolution equations for the fluid are the conservation of baryon number
\beq
\nabla_\mu (\rho u^\mu) = 0,
\eeq
the Bianchi identity (i.e.~the general relativistic equivalent of the conservation of momentum and energy)
\beq
\nabla_\mu T^{\mu\nu} = 0,
\eeq
and Maxwell's equation
\beq
\nabla_\mu (b^\mu u^\nu - b^\nu u^\mu) = 0.
\label{eq:max}
\eeq
Here, $\nabla_\mu$ is the covariant derivative defined with respect to the 4-metric $g_{\mu\nu}$. 
We use the equation of state 
\beq
P = (\Gamma -1) u \equiv \rho \Theta
\eeq
with $\Gamma$ the adiabatic index of the fluid, $\Theta = k T/(m_p c^2)$ the dimensionless temperature, and $m_p$ the proton
mass.  In the simulations, we take the adiabatic index of an ideal gas of relativistic particles, $\Gamma=4/3$.
In radiatively inefficient accretion flows, the high density regions of the disc are not relativistic, and we expect
an adiabatic index closer to $\Gamma=5/3$. However, in ideal GRMHD models,
varying $\Gamma$ from $4/3$ to $5/3$ produces only subtle changes in the flow structure~\citep{Shiokawa2013}.
When exploring the potential effects of a heat conduction and pressure anisotropy on the evolution of the disc, there are also
advantages to the use of a $\Gamma=4/3$ index. In particular, it is possible to use larger values of the conductive diffusivity and
kinematic viscosity without the model becoming acausal in high temperature regions, where the plasma would be relativistic 
(see closure relations in Sec.~\ref{sec:closure}). Practically however, as we will show, our simulations show that the important 
effect determining the amplitude of the pressure anisotropy is its saturation at the mirror and firehose instability thresholds. 
We have verified that this is the case for $\Gamma=5/3$ and $\Gamma=4/3$, and that the results discussed in this paper 
are independent of the choice of $\Gamma$, as long as the conductive diffusivity and kinematic viscosity are in the stable regime
(see  Sec.~\ref{sec:closure}).

We also need evolution equations for the heat flux and pressure anisotropy. 
These are written with respect to the rescaled variables
\beqn
\tilde{q} &=& q \left(\frac{\tau_R}{\chi \rho \Theta^2}\right)^{1/2},\label{eq:q}\\
\Delta \tilde{P} &=& \Delta P \left(\frac{\tau_R}{\nu \rho \Theta}\right)^{1/2}\label{eq:dP},
\eeqn
as
\beqn
\nabla_\mu (\tilde{q} u^\mu) &=& - \frac{\tilde{q}-\tilde{q}_0}{\tau_R} + \frac{\tilde{q}}{2} \nabla_\mu u^\mu,\\
\nabla_\mu (\Delta \tilde{P} u^\mu) &=& -\frac{\Delta \tilde{P} - \Delta \tilde{P}_0}{\tau_R} + 
\frac{\Delta\tilde{P}}{2} \nabla_\mu u^\mu.
\eeqn
That is, $\tilde{q}$ and $\Delta \tilde{P}$ are damped towards their target value $\tilde{q}_0$ and $\Delta\tilde{P}_0$
over a timescale $\tau_R$. The evolution equations follow from the requirement that the model satisfies the second law of thermodynamics, as detailed in Paper I. 
We note that Eqs~(\ref{eq:q}-\ref{eq:dP}) are obtained from Eqs. (47) and (52) of Paper I though a change of variable,
with a rescaling of the evolved variable by a factor of $\sqrt{\rho}$. This explains the difference in the form of the evolution
equations. In particular the last term in Eqs~(\ref{eq:q}-\ref{eq:dP}), which is not present in Paper I, is only due to the different 
choice of rescaled variables. We find that the variables defined in Paper I lead to larger numerical errors in low density regions.

The target heat flux and pressure anisotropy are
\beqn
q_0 &=& -\rho \chi \hat b^\mu (\nabla_\mu \Theta + \Theta u^\nu \nabla_\nu u_\mu),\\
\Delta P_0 &=& 3\rho\nu (\hat{b}^\mu \hat{b}^\nu\nabla_\mu u_\nu - \frac{1}{3} \nabla_\mu u^\mu),
\label{eq:DP0}
\eeqn
and $\tilde {q}_0$, $\Delta \tilde{P}_0$ are obtained from $q_0$, $\Delta P_0$ by the same rescaling as for 
$\tilde{q}$, $\Delta \tilde{P}$.
The target heat flux and pressure anisotropy, $q_0$ and $\Delta P_0$,
are covariant versions of the heat flux and pressure anisotropy in Braginskii's theory.
In the end, the EMHD model has three parameters: the timescale $\tau_R$ and the diffusivities $\chi$ and $\nu$, which have dimensions of length $\times$ velocity.  

\subsection{Closure Relations}
\label{sec:closure}

The EMHD model has three free parameters: $\tau_R$ , $\chi$, and $\nu$.  In non-relativistic collisional theory, $\chi = \phi c_s^2 \tau_R$ and $\nu = \psi c_s^2 \tau_R$ with $\phi$ and $\psi$ dimensionless constants of order unity and $\tau_R$ of order of the mean-free-time between collisions.  In relativistic theory we use the same form for the diffusivities, but replace the sound speed by its relativistic counterpart $c_s^2 = \Gamma P / (\rho + \Gamma u)$.

What about $\tau_R$?  The central idea of the EMHD model is to assume that $\tau_R$ is set not by the mean-free-time between Coulomb scatterings, which is long, but rather by an effective mean-free-time due to wave-particle scatterings.   In the absence of other effects a natural scale for $\tau_R$ is the dynamical time $\tau_d \approx \sqrt{r^3/(G M)}$.  

However, the relaxation timescale is likely shortened by the action of velocity-space instabilities such as the mirror and firehose, which are driven by shearing, heating, expansion, or compression of the plasma.  The single fluid model used here represents the ions in a RIAF.  Ions are unstable to the firehose instability if $\Delta P < -b^2 \equiv \Delta P_{\rm firehose}$ and to the mirror instability if
\beq
\Delta P>\frac{b^2}{2} \frac{P_\parallel}{P_\perp} = \frac{b^2}{2} \frac{3P-2\Delta P}{3P+\Delta P} \equiv \Delta P_{\rm mirror}
\eeq
(see, e.g., ~\citealt{Kunz2014}).  Once the pressure anisotropy reaches the mirror/firehose threshold the instability should grow and the pressure anisotropy should plateau.  Similarly, the heat flux should satisfy $|q| \lesssim A \rho c_s^3$ with $A\sim 1$ \citep{Cowie1977}.

The limits on pressure anisotropy and heat flux are due to an increase in the rate of pitch angle scattering by wave-particle interactions.  In the EMHD model, then, the effective collision rate should increase and $\tau_R$ should decrease near these limits.  To model this we set
\beq
\tau_R = \tau_d \times \left(f(|q|,\rho c_s^3) \times f(\Delta P,\Delta P_{\rm mirror}) \times f(-\Delta P,b^2) + f_{\rm min}\right).
\eeq
where
\beqn
f(x,x_{\rm max}) &\equiv& \frac{1}{1+e^{g(x,x_{\rm max})}}\\
g(x,x_{\rm max}) &\equiv&\frac{1}{\Delta x} \frac{x-x_{\rm max}}{x_{\rm max}}.
\eeqn
In simulations presented here we set set $\Delta x = 0.01$ and $f_{\rm min}=10^{-5}$. 
We also tested the prescription $\Delta x = 0.1$, and found no significant
changes in the results.

So far, we have focused on firehose and mirror instability.   For $\Delta P > 0$ the ion cyclotron (IC) instability is also potentially important.  The expected threshold is $\Delta P_{\rm IC} \simeq 0.35 P_{\parallel}^{1-\alpha} (b^2/2)^\alpha$ with $\alpha \approx 0.45$~\citep{Sharma2006}.
In strongly magnetized regions, $\Delta P_{\rm IC}<\Delta P_{\rm mirror}$. However, observations of the solar wind show that while the pressure anisotropy appears to saturate at $\Delta P_{\rm mirror}$ and $\Delta P_{\rm firehose}$, this is not the case at $\Delta P_{\rm IC}$ \citep{Hellinger2006}.  One possible explanation for this is that, as a resonant instability, the ion cyclotron instability is sensitive to small deviations from the bi-Maxwellian distribution function commonly assumed when evaluating velocity space instability thresholds \citep{Isenberg2013}.  In this first study we generally ignore the ion cyclotron instability when setting 
$\tau_R$, but assess its potential importance in \S \ref{sec:dP}.

We must still choose the dimensionless conductivity and viscosity $\phi$ and $\psi$. Not all choices produce stable, causal models.  In Paper I we showed that the stability of the system requires $\phi \leq (c/c_s)^4$, while causality requires $\psi \leq 0.75(c/c_s)^2$.  To satisfy these conditions for any value of the physical parameters for an adiabatic index $\Gamma=4/3$ (which implies $c_s^2<1/3c^2$), we need $\phi \leq 9$ and $\psi \leq 9/4$. In Paper I, we also showed that for $0<\phi<3$, non-linear  instabilities appear in the model if $q/u \gtrsim (3-\phi)/10$.  The model could thus be unstable even for relatively low heat fluxes when $\phi \gtrsim 3$.  In our standard model we set $\phi = 1$ and $\psi = 2/3$.  We show below that $\Delta P$ bumps up against the mirror or firehose instability thresholds over most of the computational domain, and so larger values of $\psi$ would yield very similar results. As already mentioned, these stability conditions are more constraining for a non-relativistic ideal fluid with
$\Gamma=5/3$. Analytically, and for linear perturbations around an ideal fluid state, the model is then stable if 
$\phi \leq 1.5$ and $\psi \leq 9/8$ in high temperature regions. Numerically, we find that instabilities can occur at particularly hot 
points for slightly lower values of $\phi$. To avoid this issue, when using $\Gamma=5/3$, we choose $\phi=1/2$, $\psi=1/3$.

\subsection{Numerical Methods}
\label{sec:grim}

In \S \ref{sec:eq}, we showed that the EMHD model involves 10 evolution equations 
(the time component of (\ref{eq:max})  is equivalent to the no-monopoles constraint). We recast the evolution equations in 
conservative form
\beq
\partial_t {\bf U} + \partial_i {\bf F}^i = {\bf S}
\eeq
where ${\bf U}=\sqrt{-g}(\rho u^t,T^t_\mu,B^i,\tilde{q}u^t,\Delta\tilde{P}u^t)$ is the vector of 10 ``conserved'' variables, ${\bf F}^i$ are fluxes and ${\bf S}$ are ``source terms'' which may, contrary to the usual meaning of source term, contain both time and space derivatives of the evolved variables via $q_0$ and $\Delta P_0$. 

We evolve the system using the \grim~code, a flexible code designed to be able to evolve a broad range of nonideal relativistic fluid models.  \grim~is described in detail in Paper II, together with an extensive series of tests; here we give a brief outline of the method.  

\grim~solves the EMHD equations using an implicit-explicit time stepper. All terms involving
spatial derivatives, i.e. the divergence of the fluxes and the spatial derivatives necessary to compute $q_0$ and $\Delta P_0$, are treated explicitly. All other terms are treated implicitly. Each substep of the timestepping algorithm thus involves inverting an implicit system of 10 equations {\em at each point}, which we do using the {\tt PETSc} library~\citep{petsc-efficient,petsc-web-page,petsc-user-ref}. The explicit treatment of the fluxes avoids coupling between neighboring points when solving the implicit equations, which would lead to a more computationally costly algorithm. As \grim~uses {\tt PETSc}'s automated
Jacobian assembly, it does not require the Jacobian as an input. We use a second-order in time implicit-explicit time stepper, a second-order reconstruction scheme to compute the state at the left and right side of each cell interface (MC, \cite{vanLeer1977}), and the HLL approximate Riemann solver to compute the fluxes on cell faces from the left and right states. 

As in all grid-based general relativistic simulations of a fluid, we must impose additional constraints in low-density regions. These ``floors'' are designed to avoid evolution towards unphysical states driven by accumulation of small numerical errors in low-density and/or magnetically dominated regions, and to improve the stability of the code. The floors should accomplish this without directly modifying the solution in higher density regions where the equations can be evolved reliably. In \grim~the floor conditions are $\rho \geq 0.1 b^2$, $\rho \geq10^{-3} r^{-3/2}$, $u \geq 0.002 b^2$, and $u \geq 10^{-5} r^{-3/2}$ (the maximum density in the initial conditions is $\rho = 1$; here $r$ is the Kerr-Schild or Boyer-Lindquist radius). If these inequalities are not satisfied, we increase $\rho$ and/or $u$ as needed. 

Arbitrary modification of the solution by floors seems unappealing; it would be nice to minimize invocation of the floors.  It is possible to do this by adding mass, momentum, and energy source terms inspired by the idea that high energy photons collide and create pairs in the low density, polar regions as in \cite{Moscibrodzka2012}.  We use a set of source terms that are meant to model the creation of plasma at dimensionless temperature $\Theta_W$ in the normal observer frame ($u_i=0$):
\beqn
\partial_t (\sqrt{-g}\rho u_t) &=& -10^{-6} \,\sqrt{-g} (-g^{tt})^{1/2} \,\frac{\cos^4{\theta}}{(1+r^2)^2},\\
\partial_t (-\sqrt{-g} T^t_t) &=& 10^{-6} \sqrt{-g}\frac{\cos^4{\theta}}{(1+r^2)^2} (1+ \frac{\Theta_W}{\Gamma-1}).
\eeqn
These source terms are meant to minimize invocation of the floors and should not be taken as a serious physical model (a treatment of pair production would require another fluid component with lower mean molecular weight).  We have verified that the source terms do not affect the evolution of non-polar regions by direct comparison of simulations with source terms turned on and off.  Their net effect is to significantly reduce invocation of the floors.
 
\section{Simulations}
\label{sec:ID}

In this paper we consider the axisymmetric evolution of an initially weakly magnetized, geometrically thick torus in a Kerr spacetime with $a_* = 0.9375$.

\subsection{Coordinates and Grid Structure}

The governing equations are solved on a two-dimensional uniform mesh in modified Kerr-Schild coordinates 
$t, a, b, \phi$, as in \cite{Gammie2003}.  
The coordinates are related to the usual Kerr-Schild coordinates $t, r, \theta, \phi$ by 
\beq 
r = e^{a}
\eeq
and
\beq
\theta = \pi b + \frac{1-h}{2}\sin{(2\pi b)}.
\eeq
Evidently if $\theta$ ranges over $[0,\pi]$ then $b$ ranges over $[0,1]$.  Recall that Kerr-Schild $r,\theta$ are identical to Boyer-Lindquist $r,\theta$. The model is axisymmetric (independent of $\phi$).  The coordinate parameter $h$ can be used to increase grid resolution in the disc, near the equator at $\theta = \pi/2$ or $b = 0.5$; we set $h = 0.3$.  

Our radial grid extends from $r_{\rm in}=1.321 \rS$ to $r_{\rm out}=63
\rS$.  In the equatorial plane, the horizon is located at $1.348 \rS$, the innermost stable
circular orbit (ISCO) at $r_{\rm ISCO} = 2.044 \rS$ for prograde orbits, 
and the maximum density of the initial torus is at
$r_{\rm max} = 12\rS$.  For a typical simulation that uses $280$ grid
points in $a$ there are $\sim 30$ points between the inner boundary
and the ISCO and $\sim 150$ points between the inner boundary $r_{\rm max}$.  The grid covers the entire range $\theta=[0,\pi]$.

\subsection{Initial and Boundary Conditions}

The initial conditions contain a partially pressure supported torus with constant specific entropy that follows the hydrostatic equilibrium solution of \cite{Fishbone1976}.  The initial state is uniquely determined by $r_{\rm max}$, here taken to be $12 \rS$, and the adiabat $P/\rho^\Gamma = const.$, here taken to be $0.0043$.  The density is normalized so that $\rho_{\rm max}=1$.
To this initial state we add a weak magnetic field, with azimuthal component of the vector potential given by 
\beq
A_{\phi} = A_0 \max{(\rho - 0.2,0)} \cos{(N_{\rm loops}\theta)},
\eeq
and all other components zero.  The magnetic field is $B^i = \epsilon^{ijk}
\nabla_j A_k$; here $B^i = b^i u^t - b^t u^i$ (see \citealt{Gammie2003} for
details), and $\epsilon^{ijk}$ is the antisymmetric symbol.  The constant $A_0$ sets the overall strength of the magnetic perturbation, and is chosen so that the minimum value
of $\beta = 2P/b^2$ is $\beta_{\rm min}\sim 15$.  At our standard resolution this gives about 30 points across the wavelength of the fastest growing mode of the magnetorotational instability  (MRI; \citealt{Balbus1991}) in the region in which $\beta\sim \beta_{\rm min}$.  Most of the magnetized portions of the initial disc have $\beta\sim 10^2-10^4$.

\begin{figure}
\flushleft
\includegraphics[width=1.\columnwidth]{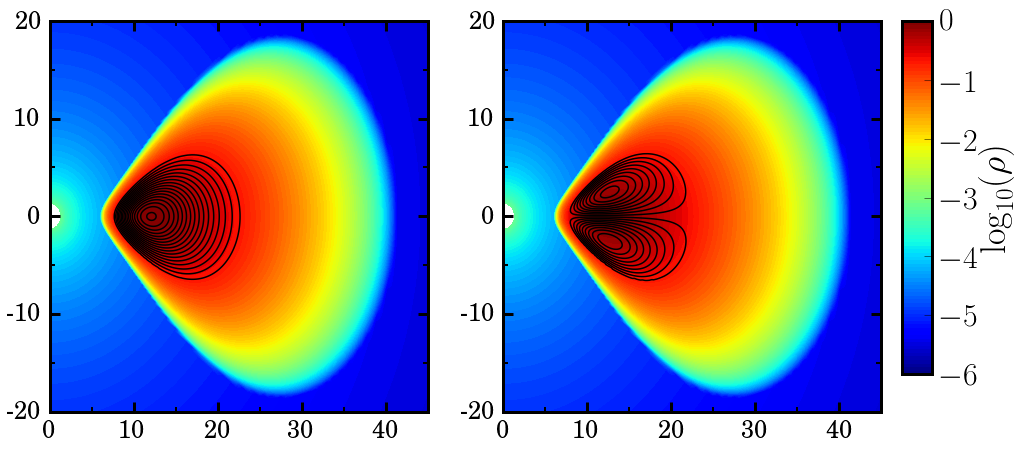}
\caption{Density (color scale) and magnetic field lines (solid lines) for the two initial conditions considered in this paper:  one magnetic field loop (left) and two magnetic field loops (right). Here and in subsequent figures, all distances are in units in which $G M = c =1$.}
\label{fig:ID}
\end{figure}

The parameter $N_{\rm loops}$ determines the number of magnetic loops in the initial disc. We consider the two configurations $N_{\rm loops}=1$ and $N_{\rm loops}=2$ (see Fig. \ref{fig:ID}) but focus primarily on the single loop case.  

The initial internal energy $u_0$ is perturbed to stimulate the growth of the MRI.  We set
\beq
u = u_0 (1 + 0.02 X)
\eeq
with $X$ a random number in the interval $[-1,1]$, different at each point in the disc. 

We also need to set boundary conditions.  At the poles we use the usual reflecting polar boundary conditions.  At the outer boundary we wish to avoid inflow, so we set $u^r=\max{(0,u^r)}$ there after each time step.  The inner boundary does not require any special treatment because we place the inner boundary inside the event horizon and this automatically enforces outflow.

Table~\ref{tab:sims} lists the simulations to be discussed in this paper.  Each initial configuration is evolved four times using four distinct models: ideal MHD, full EMHD, EMHD with conduction only, and EMHD with viscosity only.  Although the last two are not physical they provide 
insight into the relative importance of anisotropic heat conduction and viscosity.  We also perform the full EMHD one-loop evolution at three different resolutions: half the standard resolution; the standard resolution; and double the standard resolution.  Finally the one-loop configuration was run with heat flux increased by setting $\phi = 8$.  This was done because, as we will see below, the heat flux remains relatively small for $\phi = 1$ and we wanted to test the impact of a larger heat flux on the evolution of the disc. Moreover, the
exact value of $\phi$ is uncertain. Larger values of $\phi$ produces higher heat fluxes, closer to the saturated free-streaming value.
Although in principle one might expect $\phi = 8$ to be unstable and hence ill-posed (see \S \ref{sec:closure}) we did not in fact find any numerical problems with this choice of thermal conductivity, as long as $\phi$ is corrected to avoid superluminal
characteristic speeds at rare hot points appearing in the accretion flow. \footnote{Practically, we require $1+(\Gamma-1)\phi\leq c_s^{-2}$, a condition inspired by the approximate characteristic speed for the propagation of linear mode derived
in Paper I.}  We did, however, find that for $\phi=10$ the simulations became unstable.

\begin{table}
\caption{List of simulations discussed in this paper. $N_{\rm loops}$ is the number of magnetic
  loops included in the initial condition. $N_r$ is the number of radial points, $N_{\theta}$
  is the number of angular points, and $\phi$ and $\psi$ are the dimensionless constants setting
  the strength of the heat flux and pressure anisotropy, respectively.
}
\label{tab:sims}
\begin{tabular}{|c|c|c|c|c|c|}
\hline
Simulation & $N_{\rm loops}$ & $N_r$ & $N_\theta$ & $\phi$ & $\psi$ \\
\hline
Ideal MHD & 1 & 280 & 256 & 0 & 0\\
Viscosity & 1 & 280 & 256 & 0 & 2/3 \\
Conduction & 1 & 280 & 256 & 1 & 0 \\
High Conduction & 1 & 280 & 256 & 8 & 0 \\
EMHD & 1 & 280 & 256 & 1 & 2/3 \\
EMHD (Low res) & 1 & 140 & 128 & 1 & 2/3 \\
EMHD (High res)& 1 & 560 & 512 & 1 & 2/3 \\
Ideal MHD & 2 & 280 & 256 & 0 & 0\\
Viscosity & 2 & 280 & 256 & 0 & 2/3 \\
Conduction & 2 & 280 & 256 & 1 & 0 \\
EMHD & 2 & 280 & 256 & 1 & 2/3 \\
\hline
\end{tabular}
\end{table}

Table~\ref{tab:sims} does not list a number of other simulations done to test the impact of varying other numerical parameters, including atmosphere corrections, the use of lower density and internal energy floors, the addition of the wind source terms, or the use of a different adiabatic index $\Gamma=5/3$. These changes do not 
significantly alter our results and we do not discuss them further here. We also performed a simulation in which the 
pressure anisotropy saturated at the threshold of the ion cyclotron instability, which we discuss in Sec.~\ref{sec:dP}.

All models are evolved to $t = 2000 \tS \simeq 10^4 M/(10^6 M_\odot)$ sec.  In axisymmetry, disc turbulence is characterized by an initial resolution-dependent peak in turbulent field strength followed by a resolution-dependent exponential decay~\citep{Guan2008}, consistent with the anti-dynamo theorem.  Over the course of the simulation only the inner portion of the disc has time to evolve to a quasi-stationary state; the remainder of the disc is still in a transient associated with the initial conditions.  

The standard grid spacing was chosen so that the peak magnetic field strength in axisymmetry is comparable to the saturation magnetic field strength observed in fully 3D ideal MHD models~\citep{Shiokawa2012}.  Our goal in this paper is to explore the EMHD model in axisymmetry before moving on to expensive, fully 3D EMHD models.  The assumption of axisymmetry is one of the leading sources of uncertainty in our results.

\section{Results}
\label{sec:results}

\subsection{Overview}
\label{sec:imhd}

\begin{figure}
\flushleft
\includegraphics[width=1.\columnwidth]{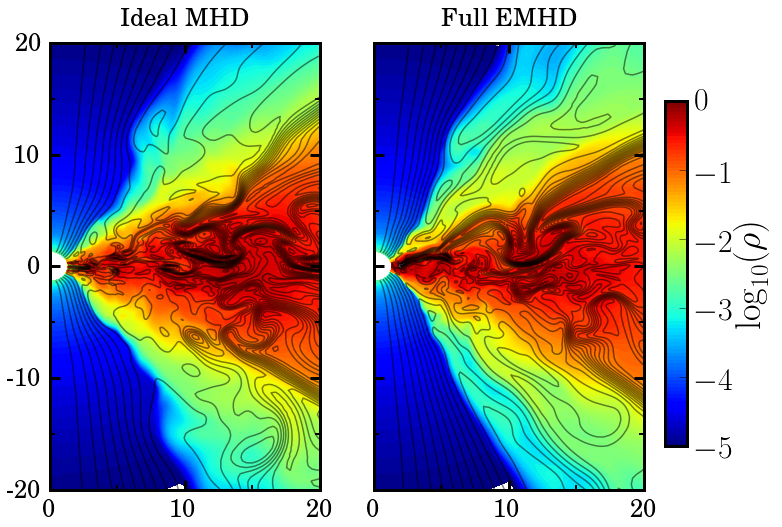}
\caption{Density (color scale) and magnetic field lines (solid lines) for the one-loop configuration at $t=1500 \tS$ for the ideal MHD ({\it left}) and full EMHD ({\it right}) models. The full EMHD and ideal MHD simulations have very similar density profiles and magnetic field
structure.}
\label{fig:RhoB}
\end{figure}

\begin{figure}
\flushleft
\includegraphics[width=1.\columnwidth]{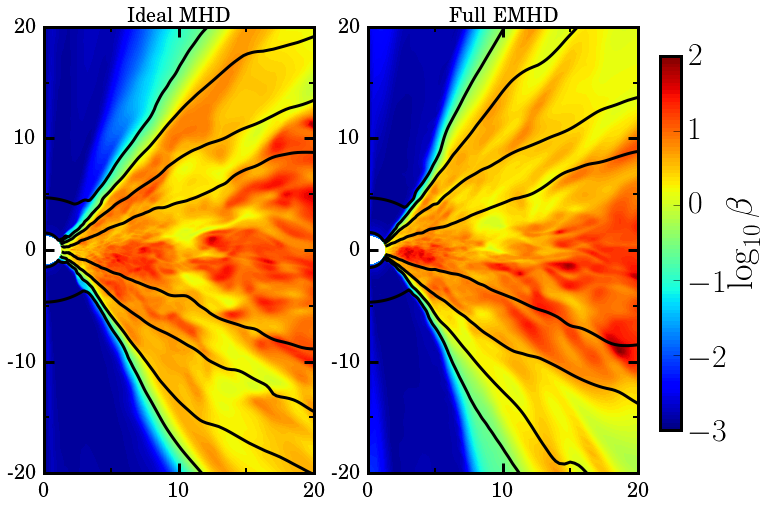}
\caption{Average plasma parameter $\beta=2\frac{\int P dt}{\int b^2 dt}$ (color scale) and density (solid lines) for the standard models in the interval $1200<t c^3/(G M) <1700$. The density contours are $\log_{10}{(\<\rho\>)}= (-4,-3,-2,-1)$. As in 
Fig.~\ref{fig:RhoB}, there are only mild differences between the EMHD and ideal MHD results.}
\label{fig:Bavg}
\end{figure}

The overall properties of the EMHD models are quite similar to the ideal MHD models. The initial state is unstable to the MRI, and then transitions to turbulence that heats the plasma and increases the magnetic field strength. leading to accretion (inflow through the event horizon) and outflows.  Fig.~\ref{fig:RhoB} compares the density and magnetic field lines at a single instant late in the evolution of both models, while Fig.~\ref{fig:Bavg} compares $\beta = 2 (\int P dt)/(\int b^2 dt)$, with the integration being done over the interval 
$1200 < t c^3/(G M) < 1700$.   The disc proper has $\beta \sim 10-100$, while the overlying corona has $\beta \sim 1-10$.  The low density region near the poles - the ``funnel'' - is very strongly magnetized, with $b^2 > \rho$.  Evidently the overall structure of the accretion flow is not strongly altered by viscosity and conduction.

The initial growth of the MRI is similar in the ideal MHD and full EMHD models.  Growth occurs most rapidly (as measured by the average velocity in the initial conditions fluid frame) in the relatively small region in which the fastest growing mode of the MRI is fully resolved.  In the standard run this is at $r \simeq 8 \rS$ in the highest latitude parts of the magnetized portion of the initial torus.  There is a short transient related to the nonequilibrium nature of the initial conditions: they are hydrodynamic but not MHD equilibria, and so they are perturbed by the addition of weak magnetic fields.  Then the average velocity grows on approximately the expected timescale $\tau_{\rm MRI} \simeq 4/(3\Omega)$\footnote{The growth rate is slightly reduced in the Kerr metric over a naive, Newtonian estimate; see \cite{Gammie2004} for more details.} ($\Omega \equiv$ disc angular velocity), until $t\sim 60 \tS$ when it saturates.\footnote{Anisotropic pressure increases the growth rate of the MRI when a toroidal field is present, but not for the purely poloidal fields initialized in our simulations (\citealt{Quataert2002,Balbus2004}).} The turbulent region then spreads through the disc, with the characteristic MRI scale $v_A/\Omega$ eventually being resolved everywhere.

\begin{figure}
\flushleft
\includegraphics[width=1.\columnwidth]{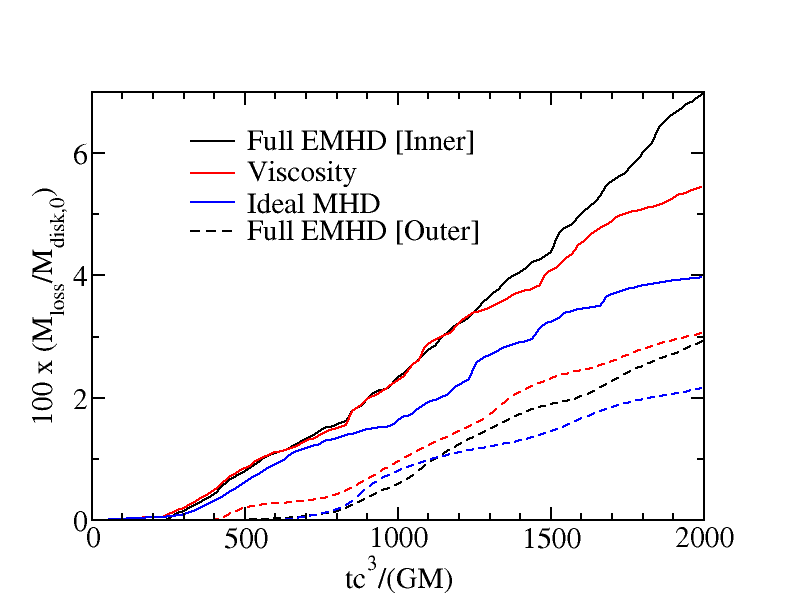}
\caption{Integrated mass lost through the inner (solid lines) and outer (dashed lines) boundaries 
for the full EMHD model, simulations with only viscosity, and ideal MHD. All simulations
start from the one-loop initial condition. The accretion rate and outflow rate are larger in the simulations
with the full EMHD model, as well as in those with only viscosity.}
\label{fig:Mdot}
\end{figure}

\begin{figure}
\flushleft
\includegraphics[width=1.\columnwidth]{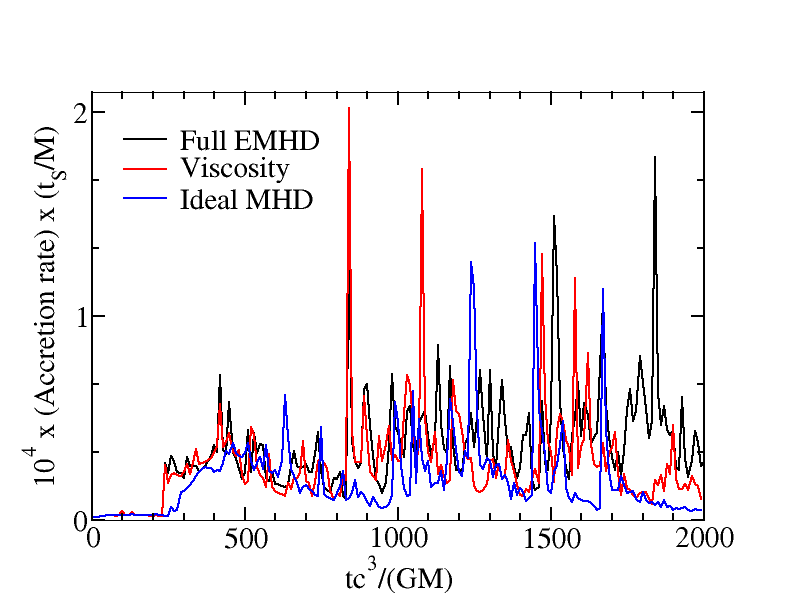}
\caption{Accretion rate onto the black hole 
for the full EMHD model, simulations with only viscosity, and ideal MHD. All simulations
start from the one-loop initial condition.}
\label{fig:AccretionRate}
\end{figure}

Nevertheless, there are significant differences between ideal MHD and full EMHD models that are visible in other time-averaged or time-integrated quantities.  First consider accretion rate and mass loss (outflow) rate.  The integrated mass lost through accretion onto the black hole (mass flux through the inner boundary) and outflows (mass flux through the outer boundary) are plotted in Fig.~\ref{fig:Mdot}, while the instantaneous mass accretion rate onto the black hole is provided in Fig.~\ref{fig:AccretionRate}. For clarity, we do not show the simulations with only conduction, which mostly track the ideal MHD case. Although there are large fluctuations, and the entire flow has not reached a quasi-equilibrium state (the accretion rate, in particular, varies significantly with the exact realization
of the turbulent flow), we note that the simulations with viscosity have larger accretion and outflow rates.   

Outflows from RIAFs are potentially important in explaining how systems like Sgr A$^*$ have low accretion rates despite an abundance of available gas at large distances from the black hole (e.g., \citealt{Blandford1999}), so the enhanced outflow rate in the full EMHD model is intriguing.  Still, the axisymmetric models are in quasi-equilibrium only relatively close to the hole.  We therefore leave a more detailed analysis of the outflow properties to future fully 3D work.  

\begin{figure}
\flushleft
\includegraphics[width=1.\columnwidth]{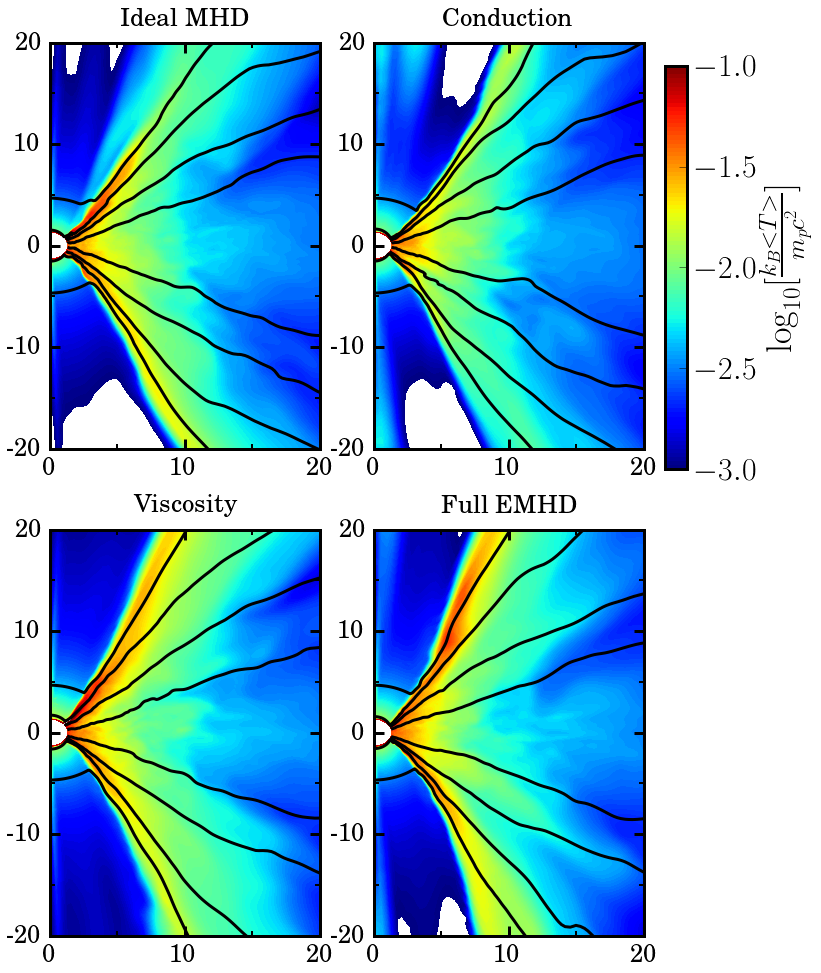}
\caption{Time averaged temperature (color scale) and density (solid lines) for the one-loop configuration in the interval  $1200<t c^3/(G M)<1700$. The density contours are for $\log_{10}{(<\rho>)}=[-4,-3,-2,-1]$.  The simulations with anisotropic viscosity have higher temperatures in the polar regions, which are associated with the $\beta \lesssim 1$ regions in Fig.~\ref{fig:Bavg} 
(see also Fig. \ref{fig:T1D}).}
\label{fig:Tavg}
\end{figure}

\begin{figure}
\flushleft
\includegraphics[width=1.\columnwidth]{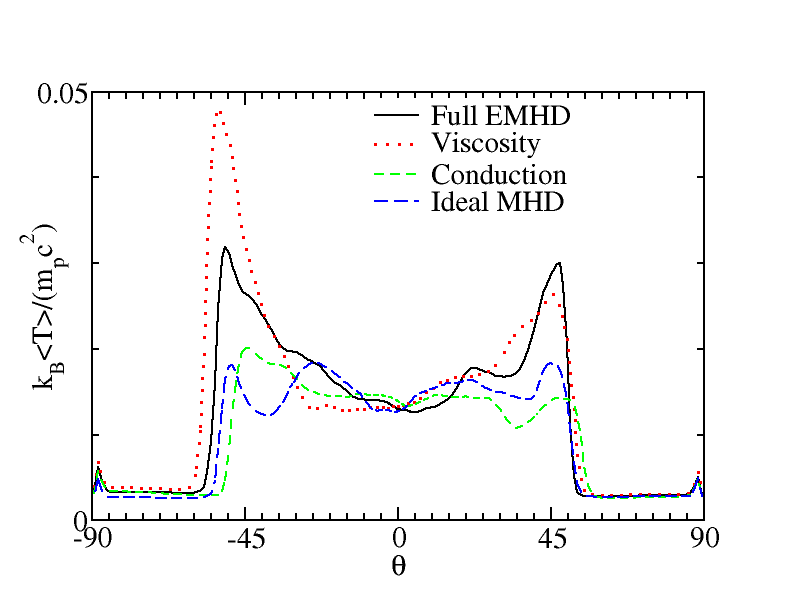}
\caption{Time and radius averaged temperature as a function of latitude for the one-loop configuration in the time interval $1200<t c^3/(G M)<1700$ and radius interval $4 < r c^2/(G M) < 7$.  The models with anisotropic viscosity have higher temperatures  by a factor of $\sim 1.5-2$ in the polar region near $\theta \sim \pm 45^\circ$. This temperature difference is even larger at large radii 
(see Fig.~\ref{fig:Tavg})}
\label{fig:T1D}
\end{figure}

The temperature structure of the full EMHD model also differs from ideal MHD models.  Fig.~\ref{fig:Tavg} show the temperature of the fluid, and density contours, averaged over $1200 < t c^3/(G M) < 1700$.  The average density is $\langle \rho \rangle \equiv (\int \rho dt)/(\int dt)$ and the average normalized temperature $\langle \Theta \rangle \equiv  (\int P dt)/(\int \rho dt)$. The most noticeable change in the full EMHD model is that the temperature is higher in the corona and funnel wall.  Conduction has a minor effect on the temperature profile, mildly smoothing temperature gradients along magnetic field lines; the increased temperature is therefore almost entirely due to viscosity.  The magnitude of the changes in $\Theta$ can more easily be assessed in Fig.~\ref{fig:T1D}, which shows the latitude dependence of the temperature in the region $4 <r c^2/(G M) <7$.    The temperature increase is even higher, a factor of 2-3, at larger radii.  Evidently the hotter corona also compresses the funnel.  

The higher ion temperature in the full EMHD models could have significant effects on the inferred observational appearance of the flow.  This will be assessed in future work that incorporates a more careful treatment of the electron thermodynamics.  

\subsection{Pressure Anisotropy}
\label{sec:dP}

The full EMHD models have an increased rate of angular momentum transport and turbulent heating as a result of viscosity, or equivalently pressure anisotropy.  The difference between ideal MHD and full EMHD is of the same order as the effect of the transport and heating in ideal MHD alone.  This is what one would naturally expect if $\Delta P \sim b^2$,  as the viscous and magnetic components of the stress-energy tensor are 
 \beqn
 T^{\mu\nu}_{\rm vis} &\sim& -b^\mu b^\nu + \frac{b^2}{3} h^{\mu\nu},\\
 T^{\mu\nu}_{\rm mag} &=& -b^\mu b^\nu + b^2 h^{\mu\nu} - \frac{b^2}{2} g^{\mu\nu},
 \eeqn
 and are thus of similar orientation and magnitude \citep{Sharma2006}. We show below that, in most of the disc, $\Delta P \approx b^2/2$.

\begin{figure*}
\includegraphics[width=2\columnwidth]{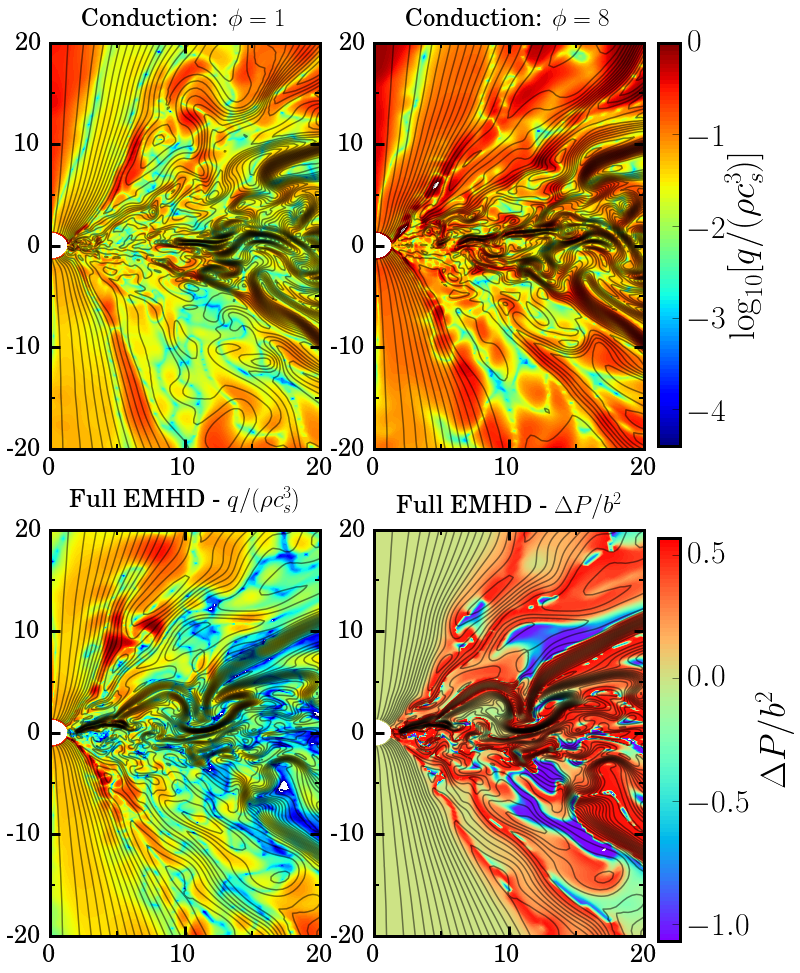}
\caption{{\bf Top:} Heat flux (color scale, normalized by $\rho c_s^3$) and magnetic field lines (solid lines) for the one-loop configuration at $t=1500\tS$, for the simulations including only heat conduction with $\phi=1$ ({\it left}) and $\phi=8$ ({\it right}). {\bf Bottom:} Heat flux ({\it left}, same color scale as top panels) and pressure anisotropy ({\it right}, color scale, normalized by $b^2$), with 
magnetic field lines for the simulation with the full EMHD model (with $\phi=1$).  For the full EMHD model, the heat flux is well below the saturated value of $q = \rho c_s^3$ and the pressure anisotropy is almost everywhere close to either the mirror ($\Delta P = b^2/2$) and firehose ($\Delta P = -b^2$) limits.  The regions close to the mirror instability threshold are associated with enhanced magnetic fields while those close to the firehose instability threshold have a lower magnetic field strength.
An animated visualization of the evolution of $\Delta P$ and $q$ in the full EHMD model is available online (http://afd-illinois.github.io/grim/gallery/).}
\label{fig:PQ1500}
\end{figure*}

The main properties of the pressure anisotropy in our simulations can be seen in Fig.~\ref{fig:PQ1500}.  The bottom right panel of Fig.~\ref{fig:PQ1500} shows $\Delta P/b^2$ at $t=1500\tS$ for the full EMHD simulation.  The simulation without heat conduction is indistinguishable.  An animated visualization of the evolution of $\Delta P$ and $q$ for the full EMHD model is also provided online~\footnote{http://afd-illinois.github.io/grim/gallery/}.  

The key result here is that, in most of the evolved domain, pressure anisotropy saturates at the mirror or firehose limits that are imposed in the model.  Even in the funnel, where $\Delta P \ll b^2/2$, the pressure anisotropy is driven close to the mirror instability limit (for $\beta \ll 1$, $\Delta P_{\rm mirror} \ll b^2/2$). This shows that the flow strongly drives $\Delta P$ up against the instability thresholds, and thus that the saturated state of the instabilities are relevant to the flow evolution.  

Furthermore, $\Delta P>0$ in nearly all regions in which the magnetic fields are large, and there is a strong correlation between field strength and $\Delta P$. As $\Delta P \sim \min{(b^2/2,P)}$, the more strongly magnetized regions are also those where the absolute magnitude of the pressure anisotropy is the strongest, and the regions with $\beta \sim 1$
are those in which the viscous stress can 
affect the evolution of the plasma the most.  In the core of the disc, $\Delta P \sim \Delta P_{\rm mirror} \sim b^2/2$ wherever the viscous and magnetic components of the stress-energy tensor are dynamically important. From these observations, we can easily explain the evolution of the global quantities studied
in the previous section: the pressure anisotropy effectively increases the magnetic stress and thus the heating by about $50\%$.

\begin{figure}
\flushleft
\includegraphics[width=1.\columnwidth]{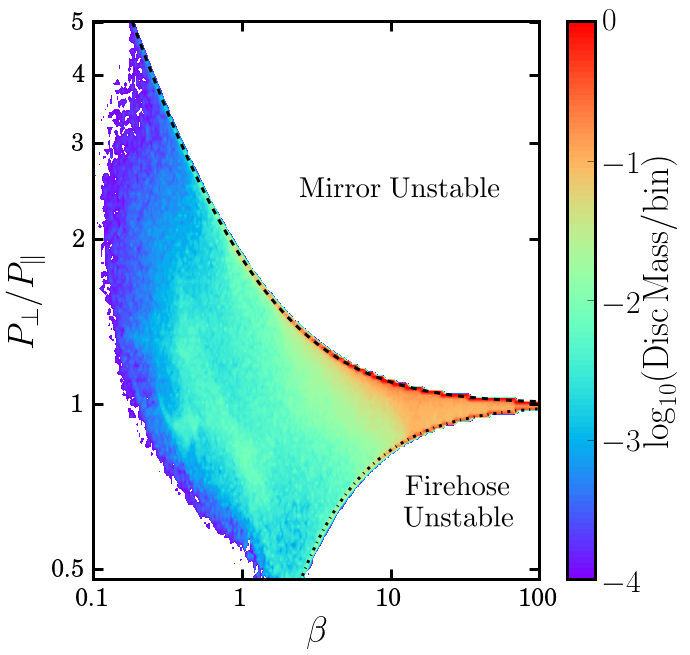}
\caption{Distribution of the pressure anisotropy $P_\perp/P_\parallel$ and plasma $\beta$
in the disc, for the full EMHD model. 
The colour scale shows the total disc mass in a given bin, summed over $51$ equally spaced snapshots of the simulation
in the time interval $1200<t c^3/(G M)<1700$ and normalized by the mass in the most populated bin. The dashed black curve shows
the saturation amplitude of the mirror instability, $\Delta P_{\rm mirror}$, and the dot-dashed black curve the saturation amplitude 
of the firehose instability, $\Delta P_{\rm firehose}$. About half of the matter has $\Delta P>0.99\Delta P_{\rm mirror}.$}
\label{fig:dPdis}
\end{figure}

\begin{figure}
\flushleft
\includegraphics[width=1.\columnwidth]{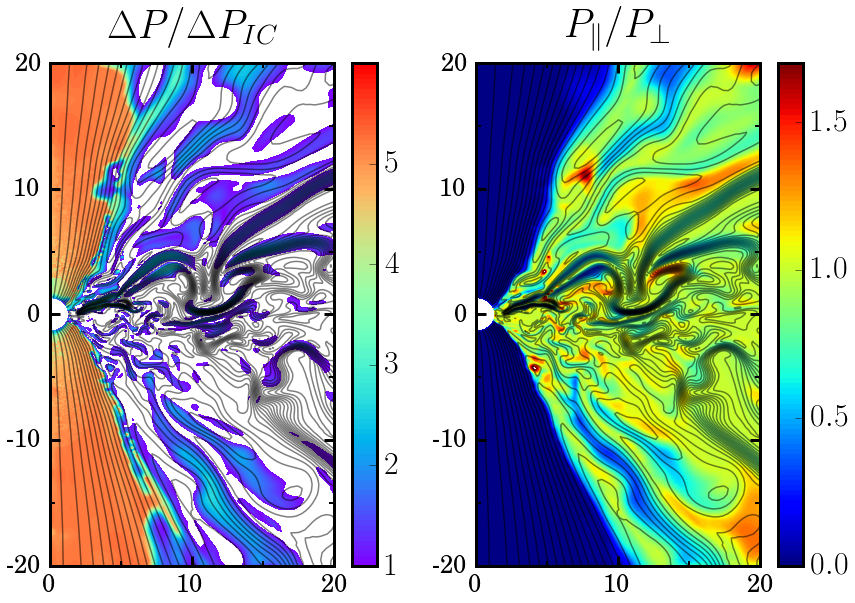}
\caption{{\it Left:} Ratio of the pressure anisotropy $\Delta P$ to the theoretical threshold $\Delta P_{\rm IC}$ to become unstable to the ion cyclotron instability, in regions where $\Delta P>\Delta P_{\rm IC}$ (white regions have $\Delta P<\Delta P_{\rm IC}$). 
We show results for the one-loop configuration at $t=1500\tS$. {\it Right:} Ratio of the pressure parallel to and orthogonal to the magnetic field lines.}
\label{fig:presscomp}
\end{figure}

Fig.~\ref{fig:dPdis} shows the distribution of the pressure anisotropy (by rest mass) in the $\beta, P_\perp/P_\parallel$ plane, similar to Figure 1 in an analysis of solar wind data  by \cite{Hellinger2006}, while Fig~\ref{fig:presscomp} (right panel) shows the spatial
distribution of $P_\parallel/P_\perp$ at $t=1500 GM/c^3$. Evidently the bulk of the fluid is at $\Delta P>0$ ($P_\perp>P_\parallel$).  The dashed lines on Fig.~\ref{fig:dPdis} show the mirror and firehose instability thresholds; a large fraction of the disc mass is at $\Delta P \approx \Delta P_{\rm mirror}$, and a smaller but significant fraction of the disc is at $\Delta P \approx \Delta P_{\rm firehose}$. In our simulations, between $1200<t c^3/(G M)<1700$, $49\%$ of the matter has $(\Delta P/\Delta P_{\rm mirror})>0.99$, and $15\%$ of the matter has $(\Delta P/\Delta P_{\rm firehose})>0.99$. In more magnetized regions, the bias towards $P_\perp>P_\parallel$ is 
stronger, but a smaller fraction of the matter is pushed against the mirror and firehose instability thresholds. The fraction of the matter close to these thresholds becomes $38\%$ (mirror) and $2\%$ (firehose) for $\beta<10$, and $9\%$ (mirror) and $0\%$ (firehose) for $\beta<1$.  The exact distribution of matter around the instability limits is controlled by the parameter $\Delta x$, which has been chosen arbitrarily but should be calibrated by PIC models. Nevertheless, it is evident that most of the plasma is driven up against the instability boundaries by the flow. 

What should we have expected for $\Delta P$?  The target pressure anisotropy is, from (\ref{eq:DP0}),
\beq
\langle \Delta P_0 \rangle = 
\langle 3\rho\nu \left(\hat{b}^\mu \hat{b}^\nu\nabla_\mu u_\nu - \frac{1}{3} \nabla_\mu u^\mu\right)\rangle \simeq 3 \rho \nu 
\langle \hat{b}^\mu \hat{b}^\nu \rangle \nabla_\mu u_\nu
\eeq
where $\langle\rangle$ indicates a time average, and in the final estimate we have assumed that density, viscosity, and velocity fluctuations are uncorrelated with magnetic field fluctuations.  For simplicity we will estimate $\langle \Delta P_0\rangle$ far from the black hole, in the nonrelativistic regime, at the disc midplane.  The Keplerian angular speed $\Omega \sim r^{-3/2}$ and taking $B_r$ and $B_\phi$ to temporarily mean the field components in elementary spherical polar coordinates (orthonormal basis, not coordinate basis), 
\beq
\langle \Delta P_0 \rangle \simeq \frac{9}{2} \rho\nu \Omega \frac{\langle -B_r B_\phi \rangle}{\langle B^2\rangle}.
\eeq
The final, magnetic term is a geometrical factor known to be $\sim 1/4$ from local, stratified models of accretion disc turbulence \citep{Guan2009}.  Evidently $\Delta P_0 > 0$ in the bulk of the disc.  

Using the closure model $\nu \sim \psi c_s^2 \tau_R$ and the nonrelativistic estimate $\rho c_s^2 \simeq \Gamma P$,
\beq
\frac{\langle \Delta P_0 \rangle}{P} \simeq \frac{9}{8} \psi \Gamma \Omega \tau_R.
\eeq
We expect mirror instability if
\beq
\frac{9}{8} \psi \Gamma (\Omega \tau_R) \gtrsim \frac{1}{\beta}
\eeq
(recall that $\beta \equiv 2P/b^2$).  For $\psi = 2/3$, $\Gamma = 4/3$, $\Omega \tau_R = 1$ we expect mirror instability if $\beta \gtrsim 1$, which is typically satisfied below about two disc scale heights.  Therefore, we should have expected mirror instability in the bulk of the disc.  

What is perhaps surprising is that the corona and funnel of the disc are also predominantly mirror-unstable. However, the sign
of $\Delta P$ in that region can be explained through a fairly simple physical argument.
The pressure anisotropy evolution depends on how $b^3/\rho^2$ varies with radius, with increasing $b^3/\rho^2$ leading to $\Delta P > 0$ and the mirror/ion cyclotron instabilities, while decreasing $b^3/\rho^2$ leads to $\Delta P < 0$ and the firehose instability
(see Appendix A of Paper I).  In the outflowing polar regions of the disc, the magnetic field is predominantly toroidal, so that $b \propto r^{-1}$; a steady outflow with $\rho \propto r^{-2}$ thus should generically lead to the mirror and ion cyclotron instabilities.   For this reason it seems likely that the prominence of instabilities associated with $\Delta P > 0$ in both the midplane and polar outflows is generic.   

We now return to the potential importance of the ion cyclotron instability. The left panel of Fig.~\ref{fig:presscomp} shows the ratio of $\Delta P$ to the threshold of the ion cyclotron instability $\Delta P_{\rm IC}$.  Regions with $\Delta P < \Delta P_{\rm IC}$, in which
the ion cyclotron instability can be self-consistently neglected, are shown in white.   Fig.~\ref{fig:presscomp} shows that the region in which the temperature is larger because of viscosity is also where $\Delta P_{IC} \lesssim \Delta P$.  It is possible that the flow structure, and in particular the heating of the outflows, would be different if $\Delta P$ saturated at $\Delta P_{\rm IC}$.  Although this is not observed in the solar wind, it would  be valuable to better understand why this is not the case, and what is the precise influence of the ion cyclotron instability on the evolution of the low collisionality plasmas in RIAFs (see~\citealt{Sironi2015} for studies along this line). As a first test, in our 2D simulations, we can force $\Delta P$ to saturate at the ion
cyclotron instability threshold $\Delta P_{\rm IC}$. Naturally, this leads to changes in the results presented in Fig.~\ref{fig:dPdis},
with highly magnetized regions now being limited to $\Delta P < \Delta P_{\rm IC}$. We also observe a mildly thicker and hotter 
corona, but not at a level at which any systematic effect could reasonably be claimed from our 2D simulations.

\subsection{Heat Conduction}
\label{sec:q}

Heat conduction can have a dramatic effect when $q \sim \rho c_s^3 \equiv$ the saturated heat flux.  In this case standard estimates show that conduction can erase temperature gradients on a timescale of order of the dynamical time.  Fig.~\ref{fig:PQ1500} shows that when conduction is included in the model but viscosity is not, $q/(\rho c_s^3)$ is typically $1\%-10\%$.   Larger $q/(\rho c_s^3)$ are preferentially present in the hot polar regions.  

\begin{figure*}
\includegraphics[width=2.\columnwidth]{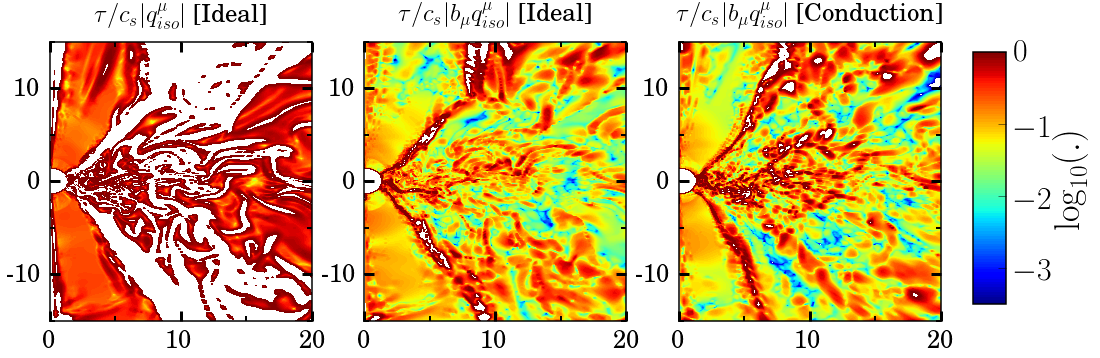}
\caption{Expected ratio of the heat flux to the saturated heat flux for isotropic conduction, with saturated regions shown in white ({\it left}) and for anisotropic conduction in the ideal MHD simulation ({\it middle}) and  in the simulation with conduction turned
on ({\it right}). This comparison shows that the suppression of the heat flux relative to the saturated value is primarily due to the small projection of the temperature gradient along magnetic field lines.}
\label{fig:GradT}
\end{figure*}

In order to understand why the heat flux is only a small fraction of the saturated value, we examine 
$q/(\rho c_s^3) \approx q_0/(\rho c_s^3) \approx (\phi \tau_R /c_s) \hat{b}^\mu q_{\mu}^{\rm iso}$,
with $q_\mu^{\rm iso}=-(\nabla_\mu \Theta + \Theta u^\nu \nabla_\nu u_\mu)$. $q_\mu^{\rm iso}$ would be the expected heat flux
if heat conduction occurred in all directions, instead of being limited to follow magnetic field lines. 
Fig.~\ref{fig:GradT} (middle and right panels) 
shows that $(\phi \tau_R /c_s) \hat{b}^\mu q_{\mu}^{\rm iso}$ is indeed typically $1\%-10\%$ in the disc. The left panel of Fig.~\ref{fig:GradT} shows the estimated $q$ {\em without} the projection of $q_\mu^{\rm iso}$
 on the magnetic field direction $\hat{b}^\mu$. We observe that $q$ then easily reaches its expected saturation value. This demonstrates that the comparatively low values of the heat flux arise in part because the temperature gradients are nearly orthogonal to the magnetic field: an isotropic prescription for the heat flux would yield $q/(\rho c_s^3) \sim 1$.  But is the misalignment of the field and temperature gradient somehow driven by the conduction itself?  The middle and right panels show the expected $q$ estimated from an ideal MHD simulation and in a conduction-only EMHD model with $\phi = 1$.  Since they are similar, we conclude that the lack of field-aligned temperature gradients is not driven by conduction.  

The simplest explanation for the misalignment between the temperature gradient and magnetic field is that the field is nearly toroidal, as in almost all simulations of magnetized turbulence in discs.  Meanwhile the temperature gradient is entirely poloidal in our axisymmetric models.  In fully three dimensional models it seems likely that the temperature gradient will remain nearly poloidal, but there is a possibility that the resulting configuration is subject to the magneto-thermal instability (MTI; \citealt{Balbus2000}).  If the MTI is vigorous enough to control the orientation of the field in some parts of the flow, then this conclusion could change.

The relative unimportance of the heat flux may depend on our somewhat arbitrary choice $\phi = 1$.  Could a larger $\phi$ change the disc structure significantly?  To test this, we evolved a model with $\phi = 8$.  Fig.~\ref{fig:PQ1500} shows that this indeed causes the heat flux to rise and get closer to saturation, at least in the more strongly magnetized regions of the disc and in the low-density outflows.  We have verified, however, that these larger $\phi$ models do not have a significantly different magnetic energy, temperature profile, or outflows, except for a modest increase in the temperature of the low-density regions at $r > 10\rS$. This might be expected because of more efficient energy transport between the inner and outer regions of the disc in regions where the radial  component of the magnetic field dominates.  The magnetic field lines late in the simulation, observable in Fig.~\ref{fig:PQ1500}, are also remarkably similar for $\phi = 1$ and $\phi = 8$.  It appears that, at least in axisymmetry, the conclusion that heat conduction has little impact on the global properties of the disc is quite robust.

\subsection{Full EMHD Model}
\label{sec:full}

The structure of the full EMHD model simulations are very similar to those obtained when pressure anisotropy is included but conduction is turned off.  This is not surprising given the results of the previous two sections.  

In fact, the coupling between the pressure anisotropy and the heat flux through the shared effective collision timescale $\tau_R$ causes the effects of heat conduction in the core of the disc to be even smaller in the full EMHD model than in the simulations evolving only the heat flux. The saturation of the pressure anisotropy, due to either the firehose or mirror instabilities, causes a strong increase in the effective collision rate, and thus (because of our closure model) a decrease in $\tau_R$. This not only stops $\Delta P$ from becoming larger than the mirror/firehose thresholds, but also significantly reduces the heat flux $q$. This is clearly visible in Fig.~\ref{fig:PQ1500}, which shows that $q$ can be an order of magnitude lower in the EMHD simulation than in the simulation modeling only heat conduction.  

Quantitatively, we find that the effective collision timescale satisfies $\tau_R \sim (\beta \Omega)^{-1}$ when $\beta \gtrsim 1$, thus reducing the heat flux by a factor of $\beta^{-1}$ in high-density regions. This dependence of the scattering rate on $\beta$ is seen in kinetic simulations of velocity space instabilities in a shearing plasma (e.g., \citealt{Kunz2014}).  It follows simply from the fact that the scattering rate required to maintain a given value of $\Delta P/P$ is $\tau_R^{-1} \sim \Omega \Delta P/P \sim \Omega \beta$.

The exception to our observation of reduced conduction in models with viscosity is at the funnel wall, where the temperature is larger in the simulations with viscosity (see Fig.~\ref{fig:PQ1500}). Larger temperature gradients in that region cause the heat flux to be more important than in the absence of viscosity, and close to its saturated value. The effect of the heat flux is thus largest in the low-density, hot regions at the disc-funnel interface.  It is as yet unclear whether this plays an important role in the radiative properties of the disc; that  will depend on the electron thermodynamics (\citealt{Moscibrodzka2014, Ressler2015}).

\begin{figure}
\flushleft
\includegraphics[width=1.\columnwidth]{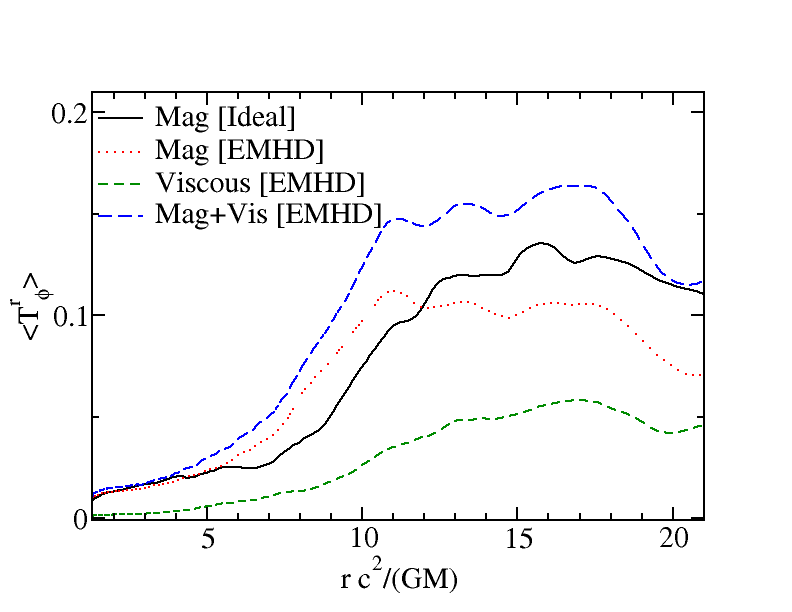}
\caption{Magnetic and viscous contributions to  $T^r_{\phi}$, a measure of the source of angular momentum transport in the disc.   Here we integrate over all polar angles and average over time within the interval $1200<t c^3/(G M)<1700$. We show results for the ideal MHD and  EMHD simulations.  Outside of the innermost stable circular orbit, the viscous shear contributes to angular momentum transport at $20\%-50\%$ of the level of the magnetic shear.}
\label{fig:Trphi}
\end{figure}

To conclude our analysis of the full EMHD model, we look at the component of the stress-energy  tensor responsible for radial transport of angular momentum, $T^r_\phi$.\footnote{$T^r_\phi$ is not gauge-invariant. However, it can still be used to extract useful information about the relative importance of viscous and magnetic transport.}  Time-averaged values of some of its components, integrated over angles, are shown in Fig.~\ref{fig:Trphi}. We use
\beqn
T^r_{\phi,{\rm mag}} &=& b^2 u^r u_\phi -b^r b_\phi\\
T^r_{\phi,{\rm vis}} &=& \Pi^r_\phi.
\eeqn
 Note that the magnetic stresses are comparable in the ideal MHD an EMHD simulations, so that the total stress is larger in the EMHD simulations because of the contribution from viscosity.

\subsection{Dependence on Initial Field Geometry}
\label{sec:moddep}

As a test of the robustness of the results discussed in the previous sections, we consider a different initial configuration in which the initial field has two loops rather than one (see Fig.~\ref{fig:ID}; $N_{\rm loops} = 2$).  We find that the conclusions reached for the one-loop configuration largely carry over to the two-loop configuration.  The increased heating and accretion due to viscous stress is confirmed in the two-loop configuration, with about $\sim 50\%$ more outflow in the full EMHD model than in the ideal MHD model.  The two-loop configuration exhibits no clear differences in mass outflow between the viscous EMHD and full EMHD models.  

\begin{figure}
\flushleft
\includegraphics[width=1.\columnwidth]{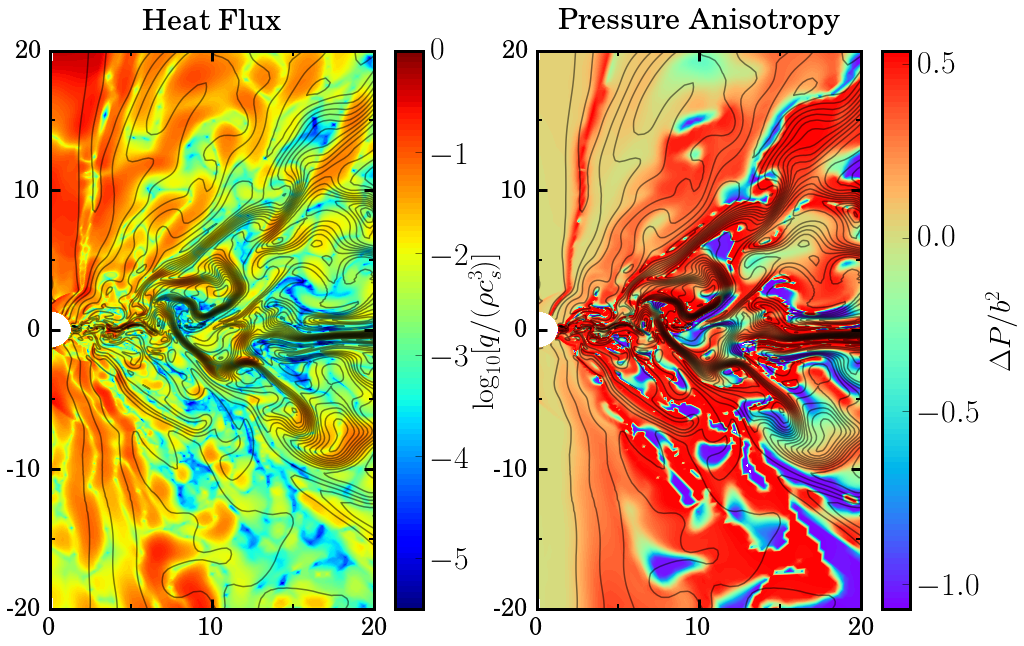}
\caption{{\it Left}: Heat flux (color scale, normalized by $\rho c_s^3$) and magnetic field lines (solid lines) for the two-loop configuration at $t=1500\tS$, for the full EMHD model. {\it Right}: Pressure anisotropy (color scale, normalized by $b^2$) and magnetic field lines (solid lines) for the same simulation at the same time. The results for the two-loop configuration are very similar to the one-loop results shown in Fig. \ref{fig:PQ1500}.}
\label{fig:PQ15002L}
\end{figure}

Fig.~\ref{fig:PQ15002L} shows that our conclusions regarding the saturation of the pressure anisotropy apply to the two-loop configuration as well: in most of the disc, and in all regions of strong magnetic fields, we have $\Delta P \approx \Delta P_{\rm mirror}$, and thus $\Delta P \sim b^2/2$ in high-density regions. The heat flux is slightly larger than in the one-loop configuration, at least in the inner disc, with $q$ at $\gtrsim 10\%$ of its saturated value when we do not evolve the pressure anisotropy. But $q$ is strongly suppressed in the full EMHD model because of the reduction in $\tau_R$ (increase in effective collisionality) associated with $\Delta P$ being pushed up against the mirror-instability boundary.

\subsection{Resolution Study}
\label{sec:res}

In three dimensional simulations of disc turbulence with an ``implicit closure'' (no explicit dissipation) a quasi-steady state can be reached with a sustained, turbulent magnetic field that is nearly independent of resolution, at least at currently accessible 
resolutions~\citep{Shiokawa2012}.   In axisymmetry this is not the case.  It is known that the saturation and eventual decay of the magnetic field is a non-convergent transient~\citep{Guan2008}.  In addition, the regions of the initial configuration in which the MRI is resolved are also resolution dependent, unless the seed magnetic field is very large or the grid spacing is smaller than used in the present study.  At our standard resolution we have only $\sim 30$ grid points across a wavelength of the fastest growing mode of the MRI in the small portion of the disc
where it is easiest to resolve (i.e., for the one-loop configuration, at the border between the magnetized and unmagnetized regions around $r=8\rS$).

\begin{figure}
\flushleft
\includegraphics[width=1.\columnwidth]{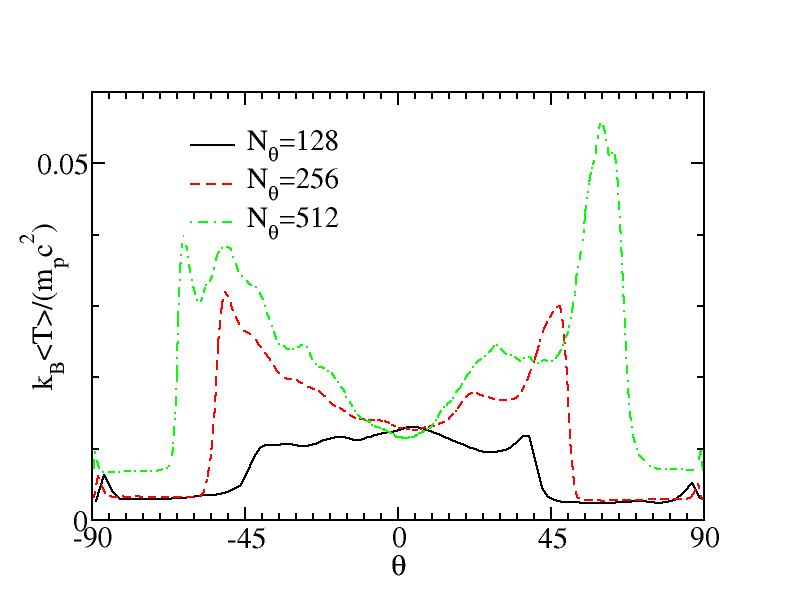}
\caption{Time and radius averaged temperature as a function of latitude for the one-loop configuration in the time interval $1200<t c^3/(G M)<1700$ and radius interval $4<r/\rS<7$, at three numerical resolutions. The higher-resolution simulations show stronger heating,
particularly in the outflow regions, and a more strongly collimated jet.}
\label{fig:TvsRes}
\end{figure}

Accordingly, the global evolution of the disc varies with resolution. More rapid accretion, stronger outflows, and a larger entropy generation are observed at high resolution, and of course the high-resolution simulation shows the growth of turbulence on smaller length scales. The most striking effect of resolution is shown on Fig.~\ref{fig:TvsRes}. The heating of the outflows is clearly not converged in 2D simulations, and increases at higher resolution. The disc and outflows cover a wider opening angle, while the highly magnetized funnel is nearly a factor of $2$ smaller at our highest resolution than at our standard resolution. This is true both for the
simulations using the full EMHD model (shown in Fig.~\ref{fig:TvsRes}), and in ideal MHD. The enhanced heating of the outflows 
in the EMHD simulations does not converge away as resolution increases, and thus non-ideal effects may affect the temperature, mass and geometry of the outflows and the opening angle of the strongly magnetized funnel; but in axisymmetry the impact of numerical
resolution on the heating of the funnel wall is comparable to the impact of non-ideal effects. Accordingly, the exact importance of that 
effect can only be determined through high-resolution 3D simulations.

Despite these differences and the limitations of axisymmetric simulations, we note that the main conclusions of this work are unaffected by numerical resolution.  The pressure anisotropy and heat conduction behave similarly at all resolutions: the pressure anisotropy saturates at $\Delta P \sim \Delta P_{\rm mirror}$ in most of the disc and in the funnel, while the heat flux remains well below its saturation value. This is as much agreement as we can expect in 2D simulations.

\section{Conclusions}
\label{sec:conclusions}

We have presented a first assessment of the differences between the global evolution of radiatively inefficient black hole accretion flows modeled as ideal magnetized fluids, and as weakly collisional plasmas. The weakly collisional model is expected to be a better representation of discs accreting well below the Eddington limit, and can thus improve our understanding of slowly accreting black holes.  This includes the majority of observed accreting black holes, e.g., the particularly important sources Sgr A$^*$ and M87.

In our extended MHD model of weakly collisional plasmas, described in Paper I, the deviations from ideal MHD are taken into account through the inclusion of a heat flux along magnetic field lines and an anisotropic viscous shear.  The latter is equivalent to the inclusion of a difference between the pressure along and orthogonal to magnetic field lines.  Our current implementation models a single fluid, best interpreted as the ions in RIAF models, since the ions generally dominate the pressure.
 
Our extended magnetohydrodynamics model is covariant, stable, and causal. In the non-relativistic limit and for slowly varying pressure anisotropies and heat fluxes, it reduces to Braginskii's theory of weakly collisional magnetized plasmas.
In our model, the magnitude of the non-ideal effects is set by the effective collision rate in the plasma. A high collision rate implies small deviations from an ideal fluid model, while a low collision rate implies larger pressure anisotropies and heat fluxes. True Coulomb collisions are very rare in the plasmas of interest in RIAFs. However, linear kinetic theory calculations and particle-in-cell simulations show that large pressure anisotropies ($\Delta P \gtrsim b^2/2$ or $\Delta P \lesssim -b^2$), can cause the growth of small-scale instabilities (e.g. the mirror, firehose, and ion cyclotron instabilities). These increase the effective collision rate in the plasma in such a way that the pressure anisotropy saturates at $\Delta P \sim b^2/2$ (or $\Delta P\sim -b^2$).  This corresponds roughly to an effective  collision timescale $\tau_R \sim (\Omega \beta)^{-1}$, at least for $\beta \gtrsim 1$.

Our simulations show that in weakly collisional accretion discs, the pressure anisotropy grows rapidly, up to the threshold for the mirror instability, in most of the disc and in the low-density, highly magnetized polar regions. The viscous stress is then comparable in magnitude and sign to the Maxwell stress, causing O(1) effects on the evolution of the disc.  This conclusion is similar to that reached by \citet{Sharma2006} and \citet{Riquelme2012} based on local shearing box fluid simulations and particle-in-cell simulations, respectively.  Our conclusions apply not just to the disc midplane, however, but also to the highly magnetized polar regions. 

In our calculations, the pressure anisotropy increases both the angular momentum transport and  the heating of the disc. The latter is particularly noticeable in the disc outflows observed at the boundary between the disc and the magnetized polar regions, and could influence the emission produced by RIAFs.

The heat flux, by comparison, has only small effects on the disc thermodynamics and structure, at least in our axisymmetric simulations.  This is in part because temperature gradients are largely orthogonal to the magnetic field lines, while particles only transport energy efficiently along magnetic field lines.  In addition, the saturation of the pressure anisotropy causes a decrease in the effective collision timescale of the plasma to $\tau_R \sim (\beta \Omega)^{-1}$ when the plasma $\beta-$parameter satisfies $\beta \gtrsim 1$. In high-density regions, 
the fluid is thus effectively more collisional, and the heat flux is suppressed by an additional factor of $\beta^{-1}$.  Whether these effects still suppress the heat flux in 3D, however, is an important open question.

An important limitation of this study is that all our simulations are axisymmetric.  Although we expect that our main conclusions are robust because they can be understood using simple physical arguments, the use of axisymmetric simulations prevents us from providing reliable predictions of the steady-state behavior of discs in the weakly collisional model.  Our conclusions are instead based on directly comparing  ideal fluid and weakly collisional simulations. In order to be able to connect these results with observations of RIAFs, 3D simulations will be required.  In addition, our current model does not treat the ions and electrons separately, while in a collisionless plasma the ions and electrons are out of thermal equilibrium. The single-fluid model used here effectively follows the properties of the ions, which is probably appropriate for determining the global properties of the disc in most cases. But the observable characteristics of the disc largely depend on the properties of the electrons. A two-fluid model like that developed in~\cite{Ressler2015} is thus necessary to directly connect simulations and observations.

Finally, it is worth noting that in regions in which $\beta \lesssim 1$, including the potentially important polar outflows, the plasma is predicted to be unstable to the ion cyclotron instability based on linear instability thresholds derived for a bi-Maxwellian distribution function.  We have not included the ion cyclotron instability in our standard model of the enhanced collisionality due to velocity space instabilities because measurements in the solar wind show that the plasma pressure anisotropy readily exceeds these nominal bounds \citep{Kasper2002,Hellinger2006}. However, if the ion cyclotron instability is in fact important in RIAF models, this could impact the pressure anisotropy and thermodynamics of the disc, particularly in the polar regions. In our 2D simulations, those differences are however weaker than the first order effect of the use of the EMHD model.

\section*{Acknowledgements}

We thank Sean Ressler, Ben Ryan, and Sasha Tchekhovskoy for discussions as well as all the members of the 
horizon collaboration, {\tt horizon.astro.illinois.edu}, for their advice and encouragement.  The horizon collaboration is supported in part by NSF .  Support for this work was provided by NASA through Einstein Postdoctoral Fellowship grant numbered PF4-150122 awarded by the Chandra X-ray Center, which is operated by the Smithsonian Astrophysical Observatory for NASA under contract NAS8-03060. MC is supported by the Illinois Distinguished Fellowship from the University of Illinois.  CFG is supported by NSF grant AST-1333612, a Simons Fellowship, and a visiting fellowship at All Souls College, Oxford.  CFG is also grateful to Oxford Astrophysics for their hospitality.  EQ is supported in part by a Simons Investigator Award from the Simons Foundation and the David and Lucile Packard Foundation, and by NSF grant AST 13-33612. 
This work was made possible by computing time granted by UCB on the Savio cluster.
This work also used the Extreme Science and Engineering Discovery
Environment (XSEDE) through allocation No. TG-AST100040, supported by NSF Grant No. ACI-1053575.




\bibliographystyle{mnras}
\bibliography{References} 

\begin{thebibliography}{}
\makeatletter
\relax
\def\mn@urlcharsother{\let\do\@makeother \do\$\do\&\do\#\do\^\do\_\do\%\do\~}
\def\mn@doi{\begingroup\mn@urlcharsother \@ifnextchar [ {\mn@doi@}
  {\mn@doi@[]}}
\def\mn@doi@[#1]#2{\def\@tempa{#1}\ifx\@tempa\@empty \href
  {http://dx.doi.org/#2} {doi:#2}\else \href {http://dx.doi.org/#2} {#1}\fi
  \endgroup}
\def\mn@eprint#1#2{\mn@eprint@#1:#2::\@nil}
\def\mn@eprint@arXiv#1{\href {http://arxiv.org/abs/#1} {{\tt arXiv:#1}}}
\def\mn@eprint@dblp#1{\href {http://dblp.uni-trier.de/rec/bibtex/#1.xml}
  {dblp:#1}}
\def\mn@eprint@#1:#2:#3:#4\@nil{\def\@tempa {#1}\def\@tempb {#2}\def\@tempc
  {#3}\ifx \@tempc \@empty \let \@tempc \@tempb \let \@tempb \@tempa \fi \ifx
  \@tempb \@empty \def\@tempb {arXiv}\fi \@ifundefined
  {mn@eprint@\@tempb}{\@tempb:\@tempc}{\expandafter \expandafter \csname
  mn@eprint@\@tempb\endcsname \expandafter{\@tempc}}}

\bibitem[\protect\citeauthoryear{Balay, Gropp, McInnes  \& Smith}{Balay
  et~al.}{1997}]{petsc-efficient}
Balay S.,  Gropp W.~D.,  McInnes L.~C.,   Smith B.~F.,  1997, in Arge E.,
  Bruaset A.~M.,   Langtangen H.~P.,  eds, Modern Software Tools in Scientific
  Computing. Birkh{\"{a}}user Press, pp 163--202

\bibitem[\protect\citeauthoryear{Balay et~al.,}{Balay
  et~al.}{2015a}]{petsc-web-page}
Balay S.,  et~al., 2015a, {PETS}c {W}eb page,
  \url{http://www.mcs.anl.gov/petsc}, \url {http://www.mcs.anl.gov/petsc}

\bibitem[\protect\citeauthoryear{Balay et~al.,}{Balay
  et~al.}{2015b}]{petsc-user-ref}
Balay S.,  et~al., 2015b, Technical Report ANL-95/11 - Revision 3.6, {PETS}c
  Users Manual, \url {http://www.mcs.anl.gov/petsc}.
Argonne National Laboratory, \url {http://www.mcs.anl.gov/petsc}

\bibitem[\protect\citeauthoryear{{Balbus}}{{Balbus}}{2000}]{Balbus2000}
{Balbus} S.~A.,  2000, \mn@doi [\apj] {10.1086/308732}, \href
  {http://adsabs.harvard.edu/abs/2000ApJ...534..420B} {534, 420}

\bibitem[\protect\citeauthoryear{Balbus}{Balbus}{2004}]{Balbus2004}
Balbus S.~A.,  2004, \mn@doi [\apj] {10.1086/424989}, 616, 857

\bibitem[\protect\citeauthoryear{{Balbus} \& {Hawley}}{{Balbus} \&
  {Hawley}}{1991}]{Balbus1991}
{Balbus} S.~A.,  {Hawley} J.~F.,  1991, \mn@doi [\apj] {10.1086/170270}, \href
  {http://adsabs.harvard.edu/abs/1991ApJ...376..214B} {376, 214}

\bibitem[\protect\citeauthoryear{{Blandford} \& {Begelman}}{{Blandford} \&
  {Begelman}}{1999}]{Blandford1999}
{Blandford} R.~D.,  {Begelman} M.~C.,  1999, \mn@doi [\mnras]
  {10.1046/j.1365-8711.1999.02358.x}, \href
  {http://adsabs.harvard.edu/abs/1999MNRAS.303L...1B} {303, L1}

\bibitem[\protect\citeauthoryear{{Braginskii}}{{Braginskii}}{1965}]{Braginskii1965}
{Braginskii} S.~I.,  1965, Reviews of Plasma Physics, \href
  {http://adsabs.harvard.edu/abs/1965RvPP....1..205B} {1, 205}

\bibitem[\protect\citeauthoryear{{Chandra}, {Gammie}, {Foucart}  \&
  {Quataert}}{{Chandra} et~al.}{2015}]{Chandra2015a}
{Chandra} M.,  {Gammie} C.~F.,  {Foucart} F.,   {Quataert} E.,  2015, preprint,
  \href {http://adsabs.harvard.edu/abs/2015arXiv150800878C} {} (\mn@eprint
  {arXiv} {1508.00878})

\bibitem[\protect\citeauthoryear{{Cowie} \& {McKee}}{{Cowie} \&
  {McKee}}{1977}]{Cowie1977}
{Cowie} L.~L.,  {McKee} C.~F.,  1977, \mn@doi [\apj] {10.1086/154911}, \href
  {http://adsabs.harvard.edu/abs/1977ApJ...211..135C} {211, 135}

\bibitem[\protect\citeauthoryear{{De Villiers}, {Hawley}  \& {Krolik}}{{De
  Villiers} et~al.}{2003}]{Devilliers2003}
{De Villiers} J.-P.,  {Hawley} J.~F.,   {Krolik} J.~H.,  2003, \mn@doi [\apj]
  {10.1086/379509}, \href {http://adsabs.harvard.edu/abs/2003ApJ...599.1238D}
  {599, 1238}

\bibitem[\protect\citeauthoryear{{Doeleman} et~al.,}{{Doeleman}
  et~al.}{2009}]{Doeleman2009}
{Doeleman} S.,  et~al., 2009, in astro2010: The Astronomy and Astrophysics
  Decadal Survey. p.~68 (\mn@eprint {arXiv} {0906.3899})

\bibitem[\protect\citeauthoryear{{Fishbone} \& {Moncrief}}{{Fishbone} \&
  {Moncrief}}{1976}]{Fishbone1976}
{Fishbone} L.~G.,  {Moncrief} V.,  1976, \mn@doi [\apj] {10.1086/154565}, \href
  {http://adsabs.harvard.edu/abs/1976ApJ...207..962F} {207, 962}

\bibitem[\protect\citeauthoryear{{Gammie}}{{Gammie}}{2004}]{Gammie2004}
{Gammie} C.~F.,  2004, \mn@doi [\apj] {10.1086/423443}, \href
  {http://adsabs.harvard.edu/abs/2004ApJ...614..309G} {614, 309}

\bibitem[\protect\citeauthoryear{{Gammie}, {McKinney}  \& {T{\'o}th}}{{Gammie}
  et~al.}{2003}]{Gammie2003}
{Gammie} C.~F.,  {McKinney} J.~C.,   {T{\'o}th} G.,  2003, \mn@doi [\apj]
  {10.1086/374594}, \href {http://adsabs.harvard.edu/abs/2003ApJ...589..444G}
  {589, 444}

\bibitem[\protect\citeauthoryear{{Guan} \& {Gammie}}{{Guan} \&
  {Gammie}}{2008}]{Guan2008}
{Guan} X.,  {Gammie} C.~F.,  2008, \mn@doi [\apjs] {10.1086/521147}, \href
  {http://adsabs.harvard.edu/abs/2008ApJS..174..145G} {174, 145}

\bibitem[\protect\citeauthoryear{{Guan}, {Gammie}, {Simon}  \&
  {Johnson}}{{Guan} et~al.}{2009}]{Guan2009}
{Guan} X.,  {Gammie} C.~F.,  {Simon} J.~B.,   {Johnson} B.~M.,  2009, \mn@doi
  [\apj] {10.1088/0004-637X/694/2/1010}, \href
  {http://adsabs.harvard.edu/abs/2009ApJ...694.1010G} {694, 1010}

\bibitem[\protect\citeauthoryear{{Hellinger} \& {Tr{\'a}vn{\'{\i}}{\v
  c}ek}}{{Hellinger} \& {Tr{\'a}vn{\'{\i}}{\v c}ek}}{2015}]{Hellinger2015}
{Hellinger} P.,  {Tr{\'a}vn{\'{\i}}{\v c}ek} P.~M.,  2015, \mn@doi [Journal of
  Plasma Physics] {10.1017/S0022377814000634}, \href
  {http://adsabs.harvard.edu/abs/2015JPlPh..81a3003H} {81, 013003}

\bibitem[\protect\citeauthoryear{{Hellinger}, {Tr{\'a}vn{\'{\i}}{\v c}ek},
  {Kasper}  \& {Lazarus}}{{Hellinger} et~al.}{2006}]{Hellinger2006}
{Hellinger} P.,  {Tr{\'a}vn{\'{\i}}{\v c}ek} P.,  {Kasper} J.~C.,   {Lazarus}
  A.~J.,  2006, \mn@doi [\grl] {10.1029/2006GL025925}, \href
  {http://adsabs.harvard.edu/abs/2006GeoRL..33.9101H} {33, 9101}

\bibitem[\protect\citeauthoryear{{Hiscock} \& {Lindblom}}{{Hiscock} \&
  {Lindblom}}{1983}]{Hiscock1983}
{Hiscock} W.~A.,  {Lindblom} L.,  1983, \mn@doi [Annals of Physics]
  {10.1016/0003-4916(83)90288-9}, \href
  {http://adsabs.harvard.edu/abs/1983AnPhy.151..466H} {151, 466}

\bibitem[\protect\citeauthoryear{{Hiscock} \& {Lindblom}}{{Hiscock} \&
  {Lindblom}}{1985}]{Hiscock1985}
{Hiscock} W.~A.,  {Lindblom} L.,  1985, \mn@doi [\prd]
  {10.1103/PhysRevD.31.725}, \href
  {http://adsabs.harvard.edu/abs/1985PhRvD..31..725H} {31, 725}

\bibitem[\protect\citeauthoryear{{Ho}}{{Ho}}{2009}]{Ho2009}
{Ho} L.~C.,  2009, \mn@doi [\apj] {10.1088/0004-637X/699/1/626}, \href
  {http://adsabs.harvard.edu/abs/2009ApJ...699..626H} {699, 626}

\bibitem[\protect\citeauthoryear{{Isenberg}, {Maruca}  \& {Kasper}}{{Isenberg}
  et~al.}{2013}]{Isenberg2013}
{Isenberg} P.~A.,  {Maruca} B.~A.,   {Kasper} J.~C.,  2013, \mn@doi [\apj]
  {10.1088/0004-637X/773/2/164}, \href
  {http://adsabs.harvard.edu/abs/2013ApJ...773..164I} {773, 164}

\bibitem[\protect\citeauthoryear{{Israel} \& {Stewart}}{{Israel} \&
  {Stewart}}{1979}]{Israel1979}
{Israel} W.,  {Stewart} J.~M.,  1979, \mn@doi [Annals of Physics]
  {10.1016/0003-4916(79)90130-1}, \href
  {http://adsabs.harvard.edu/abs/1979AnPhy.118..341I} {118, 341}

\bibitem[\protect\citeauthoryear{Kasper, Lazarus  \& Gary}{Kasper
  et~al.}{2002}]{Kasper2002}
Kasper J.~C.,  Lazarus A.~J.,   Gary S.~P.,  2002, Geophysical Research
  Letters, 29, 20

\bibitem[\protect\citeauthoryear{{Koide}, {Shibata}  \& {Kudoh}}{{Koide}
  et~al.}{1999}]{Koide1999}
{Koide} S.,  {Shibata} K.,   {Kudoh} T.,  1999, \mn@doi [\apj]
  {10.1086/307667}, \href {http://adsabs.harvard.edu/abs/1999ApJ...522..727K}
  {522, 727}

\bibitem[\protect\citeauthoryear{{Kulsrud}}{{Kulsrud}}{2004}]{Kulsrud2004}
{Kulsrud} R.~M.,  2004, {Plasma Physics for Astrophysics}

\bibitem[\protect\citeauthoryear{{Kunz}, {Schekochihin}  \& {Stone}}{{Kunz}
  et~al.}{2014}]{Kunz2014}
{Kunz} M.~W.,  {Schekochihin} A.~A.,   {Stone} J.~M.,  2014, \mn@doi [Physical
  Review Letters] {10.1103/PhysRevLett.112.205003}, \href
  {http://adsabs.harvard.edu/abs/2014PhRvL.112t5003K} {112, 205003}

\bibitem[\protect\citeauthoryear{{Mahadevan} \& {Quataert}}{{Mahadevan} \&
  {Quataert}}{1997}]{Mahadevan1997}
{Mahadevan} R.,  {Quataert} E.,  1997, \apj, \href
  {http://adsabs.harvard.edu/abs/1997ApJ...490..605M} {490, 605}

\bibitem[\protect\citeauthoryear{{McKinney} \& {Gammie}}{{McKinney} \&
  {Gammie}}{2004}]{McKinney2004}
{McKinney} J.~C.,  {Gammie} C.~F.,  2004, \mn@doi [\apj] {10.1086/422244},
  \href {http://adsabs.harvard.edu/abs/2004ApJ...611..977M} {611, 977}

\bibitem[\protect\citeauthoryear{{Mo{\'s}cibrodzka} \&
  {Gammie}}{{Mo{\'s}cibrodzka} \& {Gammie}}{2012}]{Moscibrodzka2012}
{Mo{\'s}cibrodzka} M.,  {Gammie} C.,  2012, in {Chartas} G.,  {Hamann} F.,
  {Leighly} K.~M.,  eds,  Astronomical Society of the Pacific Conference Series
  Vol. 460, AGN Winds in Charleston. p.~220

\bibitem[\protect\citeauthoryear{{Mo{\'s}cibrodzka}, {Falcke}, {Shiokawa}  \&
  {Gammie}}{{Mo{\'s}cibrodzka} et~al.}{2014}]{Moscibrodzka2014}
{Mo{\'s}cibrodzka} M.,  {Falcke} H.,  {Shiokawa} H.,   {Gammie} C.~F.,  2014,
  \mn@doi [\aap] {10.1051/0004-6361/201424358}, \href
  {http://adsabs.harvard.edu/abs/2014A%26A...570A...7M} {570, A7}

\bibitem[\protect\citeauthoryear{Quataert, Dorland  \& Hammett}{Quataert
  et~al.}{2002}]{Quataert2002}
Quataert E.,  Dorland W.,   Hammett G.~W.,  2002, \mn@doi [\apj]
  {10.1086/342174}, 577, 524

\bibitem[\protect\citeauthoryear{{Ressler}, {Tchekhovskoy}, {Quataert},
  {Chandra}  \& {Gammie}}{{Ressler} et~al.}{2015}]{Ressler2015}
{Ressler} S.~M.,  {Tchekhovskoy} A.,  {Quataert} E.,  {Chandra} M.,   {Gammie}
  C.~F.,  2015, preprint, \href
  {http://adsabs.harvard.edu/abs/2015arXiv150904717R} {} (\mn@eprint {arXiv}
  {1509.04717})

\bibitem[\protect\citeauthoryear{Riquelme, Quataert, Sharma  \&
  Spitkovsky}{Riquelme et~al.}{2012}]{Riquelme2012}
Riquelme M.~A.,  Quataert E.,  Sharma P.,   Spitkovsky A.,  2012, \mn@doi
  [\apj] {10.1088/0004-637X/755/1/50}, 755, 50

\bibitem[\protect\citeauthoryear{{Riquelme}, {Quataert}  \&
  {Verscharen}}{{Riquelme} et~al.}{2015}]{Riquelme2015}
{Riquelme} M.~A.,  {Quataert} E.,   {Verscharen} D.,  2015, \mn@doi [\apj]
  {10.1088/0004-637X/800/1/27}, \href
  {http://adsabs.harvard.edu/abs/2015ApJ...800...27R} {800, 27}

\bibitem[\protect\citeauthoryear{{Sharma}, {Hammett}, {Quataert}  \&
  {Stone}}{{Sharma} et~al.}{2006}]{Sharma2006}
{Sharma} P.,  {Hammett} G.~W.,  {Quataert} E.,   {Stone} J.~M.,  2006, \mn@doi
  [\apj] {10.1086/498405}, \href
  {http://adsabs.harvard.edu/abs/2006ApJ...637..952S} {637, 952}

\bibitem[\protect\citeauthoryear{{Shiokawa}}{{Shiokawa}}{2013}]{Shiokawa2013}
{Shiokawa} H.,  2013, PhD thesis, University of Illinois at Urbana-Champaign

\bibitem[\protect\citeauthoryear{{Shiokawa}, {Dolence}, {Gammie}  \&
  {Noble}}{{Shiokawa} et~al.}{2012}]{Shiokawa2012}
{Shiokawa} H.,  {Dolence} J.~C.,  {Gammie} C.~F.,   {Noble} S.~C.,  2012,
  \mn@doi [\apj] {10.1088/0004-637X/744/2/187}, \href
  {http://adsabs.harvard.edu/abs/2012ApJ...744..187S} {744, 187}

\bibitem[\protect\citeauthoryear{{Sironi} \& {Narayan}}{{Sironi} \&
  {Narayan}}{2015}]{Sironi2015}
{Sironi} L.,  {Narayan} R.,  2015, \mn@doi [\apj] {10.1088/0004-637X/800/2/88},
  \href {http://adsabs.harvard.edu/abs/2015ApJ...800...88S} {800, 88}

\bibitem[\protect\citeauthoryear{{Yuan} \& {Narayan}}{{Yuan} \&
  {Narayan}}{2014}]{Yuan2014}
{Yuan} F.,  {Narayan} R.,  2014, \mn@doi [\araa]
  {10.1146/annurev-astro-082812-141003}, \href
  {http://adsabs.harvard.edu/abs/2014ARA%26A..52..529Y} {52, 529}

\bibitem[\protect\citeauthoryear{{van Leer}}{{van Leer}}{1977}]{vanLeer1977}
{van Leer} B.,  1977, \mn@doi [Journal of Computational Physics]
  {10.1016/0021-9991(77)90095-X}, \href
  {http://adsabs.harvard.edu/abs/1977JCoPh..23..276V} {23, 276}

\makeatother
\end{thebibliography}

\bsp	
\label{lastpage}
\end{document}